\documentclass[12pt]{article}
\usepackage{graphicx}

\newcommand\pubnumber{SLAC--PUB--9613}
\newcommand\pubdate{December, 2002}
\newcommand\hepnumber{hep-ph/0212204}

\def\SLAC{Stanford Linear Accelerator Center\\
    Stanford University, Stanford, California 94309 USA}
\def\doeack{\footnote{Work supported by the Department of Energy,
                     contract DE--AC03--76SF00515.}}

\def\Title#1{\begin{center} {\Large #1 } \end{center}}
\def\Author#1{\begin{center}{ \sc #1} \end{center}}
\def\Address#1{\begin{center}{ \it #1} \end{center}}

\newcommand\pubblock{\rightline{\begin{tabular}{l} \pubnumber\\
         \pubdate \\ \hepnumber \end{tabular}}}
\newenvironment{Abstract}{\begin{quotation} \begin{center}
                       ABSTRACT
     \end{center}\bigskip  }{\end{quotation}}
\newenvironment{Presented}{\begin{quotation} \begin{center} 
             PRESENTED AT\end{center}\bigskip 
      \begin{center}\begin{large}}{\end{large}\end{center} \end{quotation}}

\def\beq{\begin{equation}}
\def\eeq#1{\label{#1}\end{equation}}
\def\eeqn{\end{equation}}

\newenvironment{Eqnarray}%
   {\arraycolsep 0.14em\begin{eqnarray}}{\end{eqnarray}}
\def\beqa{\begin{Eqnarray}}
\def\eeqa#1{\label{#1}\end{Eqnarray}}
\def\eeqan{\end{Eqnarray}}
\def\CR{\nonumber \\ }

\def\leqn#1{(\ref{#1})}

\let\bar=\overbar

\def\eg{{\it e.g.}}

\def\bra#1{\left\langle{ #1} \right|}
\def\ket#1{\left| {#1} \right\rangle}

\def\lsim{\mathrel{\raise.3ex\hbox{$<$\kern-.75em\lower1ex\hbox{$\sim$}}}}
\def\gsim{\mathrel{\raise.3ex\hbox{$>$\kern-.75em\lower1ex\hbox{$\sim$}}}}

\def\tr{{\mbox{\rm tr}}}
\def\half{\frac{1}{2}}
\def\thalf{\frac{3}{2}}

\def\del{\partial}
\def\Dslash{\not{\hbox{\kern-4pt $D$}}}
\def\dslash{\not{\hbox{\kern-2pt $\del$}}}

\def\ee{e^+e^-}

\def\mz{m_Z}

\def\mw{m_W}
\def\mt{m_t}

\def\mh{m_h}

\def\msb{{\bar{\ssstyle M \kern -1pt S}}}

\def\s#1{\widetilde{#1}}

\makeatletter
\def\section{\@startsection{section}{0}{\z@}{5.5ex plus .5ex minus
 1.5ex}{2.3ex plus .2ex}{\large\bf}}
\def\subsection{\@startsection{subsection}{1}{\z@}{3.5ex plus .5ex minus
 1.5ex}{1.3ex plus .2ex}{\normalsize\bf}}
\def\subsubsection{\@startsection{subsubsection}{2}{\z@}{-3.5ex plus
-1ex minus  -.2ex}{2.3ex plus .2ex}{\normalsize\sl}}

\renewcommand{\@makecaption}[2]{%
   \vskip 10pt
   \setbox\@tempboxa\hbox{\small #1: #2}
   \ifdim \wd\@tempboxa >\hsize     
       \small #1: #2\par          
     \else                        
       \hbox to\hsize{\hfil\box\@tempboxa\hfil}
   \fi}

 \def\citenum#1{{\def\@cite##1##2{##1}\cite{#1}}}
 
\newcount\@tempcntc
\def\@citex[#1]#2{\if@filesw\immediate\write\@auxout{\string\citation{#2}}\fi
  \@tempcnta\z@\@tempcntb\m@ne\def\@citea{}\@cite{\@for\@citeb:=#2\do
    {\@ifundefined
       {b@\@citeb}{\@citeo\@tempcntb\m@ne\@citea\def\@citea{,}{\bf ?}\@warning
       {Citation `\@citeb' on page \thepage \space undefined}}%
    {\setbox\z@\hbox{\global\@tempcntc0\csname b@\@citeb\endcsname\relax}%
     \ifnum\@tempcntc=\z@ \@citeo\@tempcntb\m@ne
       \@citea\def\@citea{,}\hbox{\csname b@\@citeb\endcsname}%
     \else
      \advance\@tempcntb\@ne
      \ifnum\@tempcntb=\@tempcntc
      \else\advance\@tempcntb\m@ne\@citeo
      \@tempcnta\@tempcntc\@tempcntb\@tempcntc\fi\fi}}\@citeo}{#1}}
\def\@citeo{\ifnum\@tempcnta>\@tempcntb\else\@citea\def\@citea{,}%
  \ifnum\@tempcnta=\@tempcntb\the\@tempcnta\else
  {\advance\@tempcnta\@ne\ifnum\@tempcnta=\@tempcntb \else\def\@citea{--}\fi
    \advance\@tempcnta\m@ne\the\@tempcnta\@citea\the\@tempcntb}\fi\fi}
\makeatother

\begin{document}
\begin{titlepage}
\pubblock

\vfill
\Title{Supersymmetry:  the Next Spectroscopy}
\vfill
\Author{Michael E. Peskin\doeack}
\Address{\SLAC}
\vfill
\begin{Abstract}
I describe the picture by which supersymmetry---the possible symmetry of 
Nature that converts fermions to bosons and vice versa---accounts for the next
stage of physics beyond the Standard Model.  I then survey the future 
experimental program implied by this theory, in which the spectrum of 
particles associated with supersymmetry will be determined with precision.
\end{Abstract}
\vfill
\begin{Presented}
Werner Heisenberg Centennial Symposium \\
Munich, Germany, December 5-7, 2001
\end{Presented}
\vfill
\end{titlepage}
\def\thefootnote{\fnsymbol{footnote}}
\setcounter{footnote}{0}
\tableofcontents
\newpage

\section{Introduction}

This lecture is a contribution to the celebration of the centenary of Werner
Heisenberg.  Heisenberg was one of the greatest physicists of the twentieth 
century, the man responsible for the crucial breakthrough that led to the 
final formulation of quantum mechanics.  The organizers of this Symposium have
asked me to look ahead to the physics of the twenty-first century in the 
spirit of 
Heisenberg.

This is a daunting assignment, and not just for the obvious reasons.   
The current
period in our understanding of microphysics could not be more different from 
the
period of ferment which led to the breakthrough of 1925.  Today, we have a 
`Standard Model' of strong, weak, and electromagnetic interactions that 
describes
the major facts about elementary particle interactions with great precision.  
The 
Standard Model has major problems, but these are mainly conceptual.  This 
contrasts markedly with the great periods of revolution in physics, when 
concrete
experimental data presented phenomena that could not be explained by the 
classical theory of the time or by its simple variants.

Nothing illustrates this better than the achievement of Werner Heisenberg. 
  In 
1925, classical atomic theory was beset by conceptual difficulties.  Neither 
classical mechanics nor its  direct modification by Einstein and Bohr could
explain why the atom was stable against radiation and collapse, or what 
actually
happened to an electron in the process of making a quantum transition.
Heisenberg was concerned with these issues, but his main energies went to 
problems of a very different kind.   He wanted to find the mathematical 
description
of concrete new phenomena that were emerging from the study of atomic 
spectra---the anomalous Zeeman effect, the dispersion of light in media and its
association with atomic resonances.   It is an odd and striking fact that in 
the 
fall of 1925, when  Heisenberg had already
made the breakthrough of defining and solving the quantum-mechanical
harmonic oscillator but did not yet appreciate the generality of his new
theory, he lectured at Cambridge not on his new mechanics but instead on 
the subject `Termzoologie und Zeemanbotanik'~\cite{Rechenberg}.  This 
zoological classification of the details of atomic spectra had been 
Heisenberg's main preoccupation
since the beginning of his undergraduate studies.  After the structure
of quantum mechanics had become clear, Heisenberg put the theory to the 
test against these same problems and found its success in clarifying details of
spectroscopy that were otherwise inexplicable, most notably, the spectra of
 ortho- and para-Helium~\cite{Uncertainty}.
 It was out of this struggle to find
patterns in spectroscopy that Heisenberg's quantum theory was born.

Today, some physicists talk about finding a `theory of everything' 
that will unite the interactions of microphysics with gravity and 
explain the various types of
elementary particles found in Nature. The approach is intriguing, but I am 
skeptical about it.  We
have a long way to go toward this ultimate theory. It is likely 
that it lies on the
other side of another era of experimental confusion, of crisis and resolution.
Instead of asking about final unification, we should be asking a 
different question: Where will the next
crisis in fundamental physics 
come from, and how can we help it come more rapidly?

This question is increasingly pressing as we move into the 
twenty-first century.
We have left behind long ago the era in which it is possible to 
probe new domains
of physics with a tungsten wire and a Bunsen burner. Today, probes beyond the
known realms of physics require giant accelerators, huge telescopes, massive 
detectors.  We ask governments and the public to pay for these 
endeavors, at the 
level of billions of dollars or euros.  They, in turn, ask for 
an increasingly concrete
picture of what we intend to explore and what insights we will bring back.

In this lecture, I would like to describe a path we might take to the next
corpus of data that could overturn our current physical pictures.  
Any such story
is to some extent speculative, or else completely uninteresting.  
But despite some
speculative jumps, I hope you will find this story plausible 
and even compelling.   I believe that 
there is a path to an era when we will be challenged by data to 
make a revolution
in physics, perhaps even one as profound as Heisenberg's.  The crucial element
in this path is the appearance of {\it supersymmetry} in high-energy physics.

\section{Triumphs and problems of the Standard Model}

Before explaining why supersymmetry is important, or even what it is, 
I would like
to recall the status of our current understanding of elementary particle
physics.   In 1925, there were only three elementary particles known, 
the electron, the proton, and the photon.  By the last decade of 
Heisenberg's  life, the three interactions of subatomic physics---the 
strong, weak, and electromagnetic interactions---were clearly delineated.
However, 
the first two of these were still mysterious.  For the strong 
interactions, bubble chamber experiments were turning up hundred of new
particles that needed classification.  For the weak
interactions, the property of parity violation had been discovered but its
ultimate origin remained unknown.

Today, the situation has been clarified almost completely.
 The hundreds of strong\-ly
interacting particles are now understood to be bound states of more 
elementary fermions, called `quarks'.  Three varieties of fermions with 
charge -1 are known, the electron, muon, and tau, each accompanied by a 
species of neutrino.   These `leptons' share with the quarks a very 
simple structure of couplings to heavy spin-1 bosons that accounts for 
their weak interactions.  All three interactions of elementary particle
physics, in fact, are known to be mediated by spin-1 particles.  The 
equations of motion for these particles are known to have the form
 of generalized Maxwell equations
with couplings representing the actions of a fundamental group of symmetries.
This set of equations is called a `Yang-Mills theory'~\cite{YM}; the 
spin-1 particles described are called `Yang-Mills bosons' or `gauge bosons'.
For the strong interactions, the Yang-Mills symmetry
 group is $SU(3)$; for the weak and 
electromagnetic interactions, which appear in a unified structure, the 
group is $SU(2)\times U(1)$.  The resulting structure of interacting 
quarks, leptons, and gauge bosons is called, in a somewhat 
self-deprecating
way, the `Standard Model' (SM)~\cite{HM}.  

\begin{figure}[tb]
\begin{center}
\includegraphics[width=.6\textwidth]{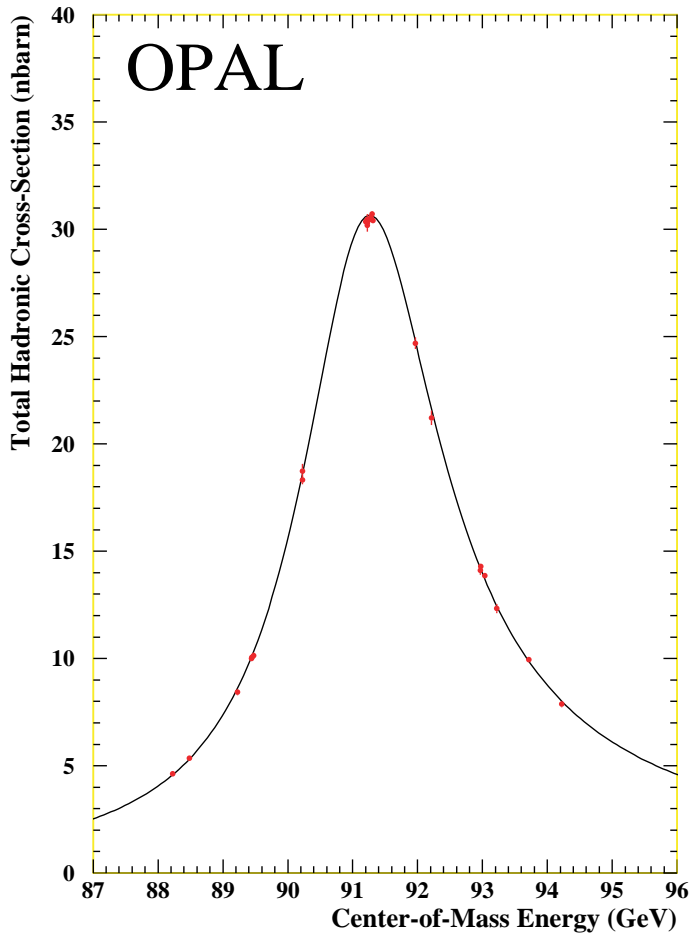}
\caption{Comparison of theory and experiment for the line-shape of the 
     $Z^0$ resonance in $\ee$ annihilation, with data from the OPAL 
       experiment~\cite{OPALZ}.}
\label{fig:Zresonance}
\end{center}
\end{figure}

  The most important result 
of high-energy physics experiments in the 1990's was the detailed confirmation 
of the predictions of the Standard Model for all three of the interactions
of elementary particle physics.  Experiments at the 
CERN collider LEP provided the centerpiece of this program, with important 
contributions coming also from SLAC, Fermilab, and elsewhere.  
Rather than give a complete review of this program, I would like to 
present just
one illustrative result.  The SM predicts that  one of the Yang-Mills bosons
mediating the weak interaction is a heavy particle called the 
$Z^0$ boson.  The $Z^0$ is a neutral particle with a mass of
about 91 GeV that can appear as a resonance in $\ee$ annihilation.  The 
resonance is a striking one: the annihilation cross section increases by a 
factor
of about 1000.  The SM predicts the width of the resonance in terms of the 
mass of the $Z^0$, the Fermi constant $G_F$, and the fine structure constant
$\alpha$.  The prediction is a sum over all species into which the $Z^0$ can 
decay, that is, over all quark and lepton species with mass less than $m_Z/2$.
In this way, the prediction invokes the basic structure of the weak 
interactions.
When quarks are produced, the decay width is enhanced by a factor 3, the number
of quantum states of the strong interaction group
$SU(3)$, and then by an extra 4\%
from strong interaction dynamics in the decay process.  Finally, the 
emission of 
photons by the electron and positron that create the $Z^0$ distorts the
resonance from a simple Breit-Wigner line-shape, causing the resonance to be 
somewhat reduced in height and more weighted to high energies.  Thus, the
complete theory of the line-shape involves detailed properties of 
all three of the 
basic interactions of microphysics.  In Fig.~\ref{fig:Zresonance}, I show the
comparison of this theory to the experimental data of the OPAL experiment
at LEP.  The agreement is extraordinary. The residual difference between
 theory
and experiment in the extracted $Z^0$ lifetime is at the level of 
parts per mil~\cite{OPALZ,allZ}.

The success of the SM in explaining this and similar data makes a strong
case for the idea that the $SU(3) \times SU(2) \times U(1)$ symmetry of the 
SM is an exact symmetry of the laws of Nature.  First of all,  we see this 
symmetry experimentally in the relations among the 
couplings of quarks and leptons to the gauge 
bosons which lead to the predictions such as that 
of Fig.~\ref{fig:Zresonance}.  Second, from a theoretical viewpoint, 
the Yang-Mills equations of motion rely on their basic symmetry being 
exact; otherwise, they 
are actually inconsistent, leading to violations of unitarity and other 
severe problems.

However, for the case of the weak interaction group $SU(2)\times U(1)$,
the symmetry is not at all manifest in the masses of elementary 
particles.   The Yang-Mills symmetry
requires that the weak interaction bosons  $W^\pm$ and $Z^0$ should
be massless like the photon.  In addition, this 
symmetry group assigns different
quantum numbers to the left-handed and right-handed spin states of quarks
and leptons.  This property is actually attractive and required when 
applied to the couplings; it 
accounts for the manner in which  the weak interactions violate parity. But it 
also forbids the appearance of quark and lepton masses.

There is  a way in which symmetries of Nature can be exact and 
also appear 
broken.  It is possible that the Hamiltonian  can have an exact symmetry
but that the ground state of this Hamiltonian might not respect this 
symmetry.  As an example, consider a magnet; the Hamiltonian describing the
spins of electrons is rotationally invariant, but in the ground state the 
spins all orient in a certain direction.  This situation is 
called `spontaneous symmetry breaking'.  Many condensed matter physics systems
exhibit spontaneous symmetry breaking, including magnets, binary 
alloys (for which
the symmetry is the lattice translation), and superfluids and superconductors
(for the symmetry is the phase rotation symmetry of the atomic or electron
wavefunction).  In each case, some aspect of the atomic interactions 
causes a macroscopic degree of freedom to pick a direction with respect to 
the symmetry operation and sit down in such as way as to  hold
 that orientation uniformly 
throughout the material.

We could imagine that the Yang-Mills symmetry of the weak interactions is
spontaneously broken.  But then there is a question: What entity and what
physics are responsible for choosing the orientation uniformly throughout
space.  In the simplest realization of the SM, we postulate a new scalar
field,  called the 
`Higgs field' $\varphi$, and give it
the responsiblity
for this spontaneous symmetry breaking.
Very little is known about the Higgs field 
from experiment.
The success of the SM brings this question into tight focus:  What 
is this Higgs 
field?  Why does it appear in Nature?  Why does its energetics favor 
symmetry-breaking and orientation?

The mystery of the nature of the Higgs field is the most compelling
single problem in elementary particle physics today.   It is not 
unreasonable to create a model of new interactions of elementary particles
simply to address this question.  But there are other mysterious 
aspects of the SM and microphysics, and it would be good if a model that
explains the Higgs field also has something to say about these.
For me, the most interesting of these properties are the following:
\begin{itemize}
\item The heaviest particle of the SM is the top quark, with a mass 
much heavier
than the $W$ boson:  $\mt/\mw = 2.1$~\cite{topmass}.
\item The Higgs boson must not only exist, but it is required by the 
constraint of
the precision electroweak data to be light~\cite{LEPHiggs}
\beq
 \mh < 193\ \mbox{ GeV}\ . \qquad\mbox{(95\%\ CL)}
\eeq{Hmassconstr}
 It is possible
that the Higgs boson was observed in the last year of operation of LEP, at a
mass of 115 GeV~\cite{KadoTully}.
\item The precision experiments give quite definite values for the three gauge
coupling constants of the SM.  Writing $\alpha_i = g_i^2/4\pi$ with 
$g_i = g'_1$
for $U(1)$, $g_2$ for $SU(2)$, $g_3$ for $SU(3)$, we have found that
\beq
  \alpha_1^\prime= 1/98.4 \ , \qquad     \alpha_2= 1/29.6 \ , \qquad 
    \alpha_3= 1/8.5 \ ,
\eeq{couplings}
with errors of 2\% for the strong interaction coupling $\alpha_3$ and 
of 0.1\%  for the electroweak couplings~\cite{ErlerL}.
\item As explained in Michael Turner's lecture at this symposium, 
ordinary matter is far from being the dominant form of energy in 
the universe.  In units where the 
energy density in a flat universe is $\Omega_0 
\sim  3$ GeV/m$^3$, about 30\%
is composed of `dark matter', a 
heavy, non-luminous, non-baryonic form of matter.
And almost 70\% is composed of 
`dark energy', energy of the vacuum or of a 
new field which obtains a vacuum expectation value~\cite{Turner}.
\end{itemize}
A theory that supercedes the SM should have a place for these phenomena.

\section{Supersymmetry}

The search for a framework in which to build a theory beyond the 
SM brings us to 
supersymmetry.  Supersymmetry is a mathematical idea of a means to generalize
quantum field theory.  It was introduced in the early 1970's by Golfand and
Likhtman \cite{GL}, Volkov and Akulov \cite{VA} and Wess and Zumino \cite{WZ}.
The last of these papers, which introduced the linear representations of the 
symmetry on fields, opened a floodgate to  theoretical developments.  
In this lecture, I will
explain in the simplest terms what supersymmetry is, and then I will pursue its
implications in a way that will link with the questions of the previous 
section.
Broader reviews of supersymmetry can be found 
in many articles and books, including \cite{Nilles,WessBagger,Martin}.

Formally, a supersymmetry is a symmetry of  a quantum system
 that converts fermions to boson and bosons to fermions.
\beq
     [ Q_\alpha, H] = 0 \qquad  Q_\alpha\ket{b} = \ket{f} \qquad 
              Q_\alpha\ket{f} = \ket{b}\ .
\eeq{SUSYdef}
In relativistic quantum field theory, bosons carry integer spin and 
fermions carry
half-integer spin, so $Q_\alpha$ must have half-integer spin.  The simplest 
case is spin-$\half$.  The assumption that there exists a 
spin-$\half$ charge
that commutes with $H$ seems innocuous, but it is not.

To see this, consider the object  $\{ Q_\alpha, Q_\alpha^\dagger\}$. 
This quantity
commutes with $H$.  It carries two spinor indices; under the Lorentz
 group, it is
a component of 
a four-vector.  And, it is positive if $Q_\alpha$ is nontrivial.  
To see this, note
that
\beq
\bra{\psi}\{ Q_\alpha, Q_\alpha^\dagger\}\ket{\psi} 
= \| Q_\alpha\ket{\psi}\|^2
      + \| Q_\alpha^\dagger\ket{\psi}\|^2
\eeq{positivity}
The presence of a supersymmetry thus implies the presence of a conserved 
vector charge.  But this is a problem.  Lorentz invariance
and energy-mo\-men\-tum conservation already severely 
restricts the form of two-particle scattering amplitudes. The scattering
amplitude for a fixed initial state is a function of only one 
continuous variable,
the center-of-mass scattering angle.   If there is an additional 
conserved charge
that transforms as a vector under Lorentz transformations, there are too many
conditions for the scattering amplitude to be nonzero except at 
some discrete angles.  In quantum field theory,
the scattering amplitude must be analytic in the momentum transfer, so 
in such a case it can only be zero at all angles.  A rigorous
proof of this statement, applicable also to any conserved charge of (integer)
higher spin, has been given by Coleman and Mandula \cite{ColemanMandula}.

Only one possibility evades the theorem:  We must identify the conserved vector
charge with the known conserved energy-momentum.  That is,
\beq
\{ Q_\alpha, Q_\beta^\dagger\}  =   2 \gamma_{\alpha\beta}^\mu P_\mu\ .
\eeq{identifyP}
Let me put it more bluntly:  If a nontrivial relativistic quantum field theory 
contains a  supersymmetry charge $Q_\alpha$, the square of this charge is
{\it the energy-momentum of everything}.   If  $Q_\alpha$ is to be an exact
symmetry of Nature, it cannot be restricted to some small part of the equations
of motion.   $Q_\alpha$ must act on every particle.

It follows from this that, in a supersymmetric theory, 
every particle must have a 
partner of same rest energy or mass and the opposite statistics. 
 If there is a 
photon with spin 1, there must be a `photino' ($\widetilde \gamma$) 
with spin $\half$.  If there is a  $W^+$ boson, there must be a spin-$\half$
$\widetilde w^+$.  We have already noted 
that, in the SM, the left- and right-handed components of quark and lepton
fields have different $SU(2)\times U(1)$ quantum numbers.  This means that the
basic fields of a supersymmetry SM should include separate spin-0 fields
$\widetilde e_L$, $\widetilde e_R$, for example, or $\widetilde u_L$, 
$\widetilde u_R$.  In the following, I will follow the common terminology
by referring to the partners of 
Yang-Mills bosons as `gauginos'---`photino', `wino', `zino', `gluino'---and
to the partners of quarks and leptons as `sfermions'--`squarks', `sleptons',
`selectrons', {\it etc}.

One known fact about sfermions is that they do not exist with masses
equal to the masses of their partners.  There is no scalar particle of 
charge -1 with the mass of the electron, and their is no scalar particle
coupling to the $SU(3)$ gauge bosons with the mass of the $u$ quark.
Such particles might exist with higher masses, but this would require 
that supersymmetry is not an exact symmetry.
It is possible, however, that  supersymmetry, like the 
$SU(2)\times U(1)$ symmetry of the SM, is a spontaneously broken symmetry,
an exact symmetry of the equations of motion that does not 
lead to a symmetrical
vacuum configuration.  In that case, the supersymmetry partners of the quarks,
leptons, and gauge bosons could well be heavier than the  familiar
SM particles, but they must exist at mass values that we might
eventually reach in our experiments.

If supersymmetry acts on all fields in Nature, it must also act on the 
gravitational
field.  Indeed, a supersymmetric theory that contains gravity must also contain
a spin-$\thalf$ partner of the graviton.  Beginning with an apparently 
innocent 
assumption, we have learned that we must change the basic structural equations
of space-time.

\begin{figure}[tb]
\begin{center}
\includegraphics[width=.3\textwidth]{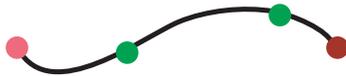}
\caption{A string is a particle which is also a one-dimension quantum system.}
\label{fig:string}
\end{center}
\end{figure}

There is another way of understanding the universal character of supersymmetry
that opens another set of connections.  Supersymmetry was originally 
discovered as a property of {\it string theory}, an idea that generalizes 
quantum
field theory by modellng particles as one-dimensional extended objects 
embedded in space-time.  The embedding is represented by a set of functions
$X^\mu(\sigma)$, where $\sigma$ is a coordinate along the string.
  Neveu, Schwarz, and Ramond~\cite{NS,Ramond} found 
that 
certain difficulties of this theory are ameliorated by adding to the string 
Hamiltonian a set of 
fermionic coordinates  $\Psi^\mu(\sigma)$.  
 (See 
Fig.~\ref{fig:string}.) The resulting quantum theory
of fields on the string has a supersymmetry, and the theory also naturally
leads to a supersymmetric theory of particles in space-time~\cite{GSO}.
The mathematical structure is that of a string moving in  a `superspace' with 
both bosonic and fermionic coordinates.  This structure becomes a part of the 
description of space-time and influences all particles that move in it.  
String theory is described in some detail in Joseph Polchinski's lecture
at this symposium~\cite{Polch}.

String theory is often described as the `theory of everything'.  
While that statement
lacks definite experimental support, string theory is a 
mathematical framework that 
successfully incorporates gravity into relativistic quantum theory. 
It is, in fact, the only known framework in which the
weak-coupling perturbation theory for gravity is well-defined
to all orders.  String theory 
also contains interesting ideas for
 how gravity fits together with the elementary
microscopic interactions.  We will find some inspiration from these ideas at a 
later point in the lecture.

\section{Supersymmetry as the Successor to the Standard Model}

I have described supersymmetry as a mathematical refinement of quantum 
field theory.  From this point of view, it is surprising that supersymmetry
can address the questions about microscopic physics that we posed in 
Section 2.  In fact, a construction based on adding supersymmetry 
straightforwardly to the SM is dramatically 
successful in resolving those questions.  This is not the only 
possible picture, but
it is, at this moment, the one which is most complete and compelling.
In this lecture, I will describe only the approach to the questions of the SM
based on supersymmetry.  For a look at the variety of other proposed 
models of $SU(2)\times U(1)$ symmetry breaking, 
see~\cite{MYCERN,EllisRev,Schmaltz}.
In only a few years---at the latest, when the Large Hadron Collider (LHC)
begins operation at CERN---we
will know whether this model or one of its competitors is correct.

Consider, then, the supersymmetric extension of the SM.  For each boson
field in the model, we add a fermion with the same quantum numbers.  For 
each fermion, we add a boson.  The interactions of these new fields are 
dictated by supersymmetry.  To this, we must add mass terms that make the
new particles heavy and other interactions that might be induced by spontaneous
supersymmetry breaking.  (These mass terms will have only a minor 
effect in this
section, but they will become significant later.)  Let us see what 
consequences
this model has for the problems discussed in Section 2.

\subsection{Higgs field}

Consider first the question of the nature of the Higgs field, its 
origin and the 
reason for its instability to spontaneous symmetry breaking.  
Within the Standard
Model, the Higgs field is anomalous.  It is the only scalar 
particle and the only 
particle that can acquire a mass without spontaneous symmetry breaking.

\begin{figure}[tb]
\begin{center}
\includegraphics[width=.5\textwidth]{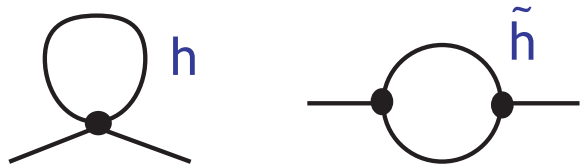}
\caption{Feynman diagrams contributing to the first loop correction to 
the Higgs boson mass.}
\label{fig:Higgsmass}
\end{center}
\end{figure}

At a deeper level, these curiosities of the Higgs boson
 turn into serious conceptual problems.
The Feynman diagrams that give higher-order corrections to the Higgs boson 
mass  are ultraviolet-divergent.   As an example, consider the first
diagram in Fig.~\ref{fig:Higgsmass}, in which the Higgs boson interacts
with its own quantum fluctuations through its nonlinear interaction.
Evaluating this contribution for momenta of the virtual Higgs boson
running up to a scale $\Lambda$, we find
\beq
       \mh^2 = \mh^2({\rm bare}) + {\lambda\over 8\pi} \Lambda^2  + \cdots \ ,
\eeq{Higgsmasseq}
where $\lambda$ is the Higgs field nonlinear coupling.  If the SM is
valid up to the scale where quantum gravity effects become important, this 
equation should be the correct first approximation to the Higgs boson mass
for the value
$\Lambda \sim 10^{19}$ GeV.   We have already noted that $\mh$ itself
is of order 100 GeV.  Thus, in the SM, the bare Higgs mass parameter and the
higher-order corrections must cancel in the first 36 decimal places.  

This 
type of
delicate cancellation is familiar from the theory of second-order phase 
transitions
in condensed matter systems.  Anyone who has experimented 
on a liquid-gas critical point knows that the temperature and pressure must
 be 
delicately adjusted to see the characteristic phenomena of the critical 
point, for
example, the critical opalescence that results from density fluctuations on a 
scale much larger than the atomic size.  In a fundamental theory of Nature, we
would like this delicate adjustment to happen automatically, not as some whim
of the underlying parameters.  

\begin{figure}[tb]
\begin{center}
\includegraphics[width=.5\textwidth]{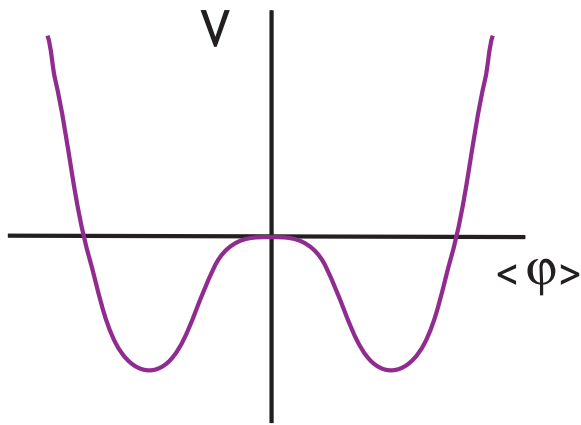}
\caption{General form of a Higgs potential unstable to symmetry breaking.}
\label{fig:Higgspotential}
\end{center}
\end{figure}
Further, if $\mh^2$ is the result of such a 
cancellation, it is an accident that the parameter should be negative
rather than positive, 
giving an
unstable potential such as that shown in Fig.~\ref{fig:Higgspotential}. 
But if we cannot predict the sign of $\mh^2$, we cannot explain why
the electroweak gauge symmetry should be broken.

Supersymmetry repairs these problems one after another.  First of all, 
supersymmetry gives a {\it raison d'etre} for the appearance of a scalar 
field.   In a 
supersymmetric generalization of the SM, there are many scalar fields, since
every quark and lepton must have a spin-0 partner.
Potentially,
any of these fields could acquire a vacuum expectation value and break the
symmetries of the model.  So we must ask why only the Higgs field has an 
instability.  I will address this problem in a moment.

Next, we should analyze the problem of large higher-order 
corrections to 
the Higgs boson mass.  In the supersymmetric SM, the calculation of $\mh^2$
has additional contributions.  One of these is shown as the second diagram 
in Fig.~\ref{fig:Higgsmass}: In addition to loop diagrams containing Higgs 
bosons, supersymmetry requires diagrams containing the 
 spin-$\half$ partners of Higgs bosons. 
In a theory with unbroken supersymmetry, the terms in these diagrams 
proportional to $\Lambda^2$ precisely cancel.
  This is a natural consequence
 of supersymmetry:  In quantum field theory, chiral symmetry requires that the
higher-order corrections to the mass $m_f$ of a fermion are of the form
\beq
         m_f =  m_f(bare) +  a_f{\lambda\over 4 \pi} m_f
                      \log {\Lambda^2\over m_f^2}\ ,
\eeq{mfcorrection}
where $a$ is a numerical constant.  The radiative correction 
to the electron mass
in quantum electrodynamics,
for example, has this form.  By supersymmetry, the bosonic partner of 
this fermion must have the same mass corrections.  In a theory with 
spontaneous supersymmetry breaking, the boson and fermion mass corrections
need not be equal.  However, since spontaneous symmetry breaking is a 
property of the lowest-energy state of the theory, 
 it cannot affect the structure deep in the 
ultraviolet. Then the boson mass is still corrected only by terms of the form
\beq
         m^2 =  m^2(bare) +  a  {\lambda\over 4 \pi} m^2\log 
{\Lambda^2\over m^2}\ ,
\eeq{mcorrection}

Having established the validity of the form \leqn{mcorrection}, 
we might next ask
what is the value of the coefficient $a$.  This question is more 
significant than it 
might appear at first sight.  If $a$ is negative, the corrected 
$m^2$ is negative if
the bare value of $m^2$ is sufficiently smaller than $\Lambda^2$.  If $a$ is 
negative and the bare value of $m^2$
 is computable from a theory of spontaneous
supersymmetry breaking, we can build
 a quantitative theory of $SU(2)\times U(1)$ symmetry-breaking.
In the supersymmetric generalization of the SM, there are a variety of 
contributions to $a$ coming from the various quarks, leptons, and gauge bosons 
that can contribute to loop corrections to the Higgs potential.  However, 
if the 
top quark is heavy, it must couple especially strongly to the Higgs field.  
Then this contribution to \leqn{mcorrection}---the contribution with
top quarks and their scalar partners in the loop---is the dominant one.  
That contribution is negative,
by explicit calculation, and drives the instability of the Higgs potential to 
spontaneous symmetry breaking.  It turns out also that, 
for a large region of the parameter space,
the Higgs is the only unstable mode among the many scalar
fields of the theory.

Thus, supersymmetry gives an origin for the Higgs field. It also explains its 
instability to spontaneous symmetry breaking by relating this to the observed 
large mass of the top quark.

\subsection{Coupling constants}

In \leqn{couplings}, I have reported the values of the three elementary 
coupling constants of the SM as determined by the recent precision 
experiments.  Supersymmetry gives the relation among these values.

In quantum field theory, coupling constants are not absolute.  They vary as a 
function of the distance scale on which they are measured, according to the
properties of the interaction.  Again, the behavour of 
quantum electrodynamics (QED)
provides a reference point.  In QED, electron-positron pairs can appear and
disappear in the vacuum as quantum fluctuations.  These evanescent pairs
give the vacuum state of QED dielectric properties.  As one approaches 
a charged particle
very closely, coming inside the polarization cloud, one sees a 
stronger charge.  Since electron-positron production in the vacuum occurs on
all length scales (smaller than the electron Compton wavelength),  
the strength of
a charge in QED appears to increase systematically on a logarithmic scale of
distance.  More precisely, the values of $\alpha = e^2/4\pi$ at two large mass
scales are related by  
\beq
   \alpha^{-1}(M) =   \alpha^{-1}(M_*) -  
                  {b\over 2 \pi} \log {M\over M_*} + \cdots\  ,
\eeq{running}
where $b$ is a constant that can be straightforwardly computed using Feynman
diagrams.  The sign $b< 0$ corresponds to charge screening by vacuum
polarization.

Similar considerations apply to the three coupling constants of the SM.  All
three couplings change slowly, as a logarithmic function of the mass or 
distance
scale.  In a non-Abelian gauge theory, there is a new physical effect that
 allows
the coefficient $b$ to be positive, so that the value of $g$ or $\alpha$ 
decreases at very short distances or large momenta. In general, the value of 
$b$ is a sum over the contributions of all particles that couple to the
bosons of the gauge theory, including quarks, leptons, Higgs bosons, and,
in the non-Abelian case, the gauge bosons themselves.

It is attractive to speculate that all three of the intereractions of the 
SM arise from a single, unified, non-Abelian
gauge symmetry, called the `grand unification' symmetry group.
  The splitting of the three interactions would result from the
spontaneous breaking of the grand unification group to the SM gauge group
$SU(3)\times SU(2)\times U(1)$.   The values of the three coupling constants
must be equal at the mass scale of 
this symmetry-breaking, but then, by the effects
just explained, they will differ at larger distance scales.  The coupling 
constant of the
$U(1)$ factor, $\alpha_1$, will be the smallest; the coupling of the largest 
non-Abelian group, the $SU(3)$ coupling $\alpha_3$, will be the largest. 
 This is 
just the pattern actually seen in \leqn{couplings}.  

We must now investigate whether this picture gives a 
quantitative explanation of
the magnitudes of the three couplings.  Before we 
begin, there is one subtlety
to take care of.  The normalization of the coupling constant of a non-Abelian
group is unambiguous, but, for an Abelian group, this 
normalization is a matter 
of convention.  The coupling
\beq
         \alpha_1 =  {5\over 3} \alpha'_1
\eeq{alphaonedef}
is correctly normalized so that it equals $\alpha_2$ and $\alpha_3$ at the 
scale of grand unification symmetry breaking in the case of grand unification
groups  $SU(5)$, $SO(10)$, and $E_6$, the groups that are attractive 
candidates for the unification symmetry because their simplest representations 
reproduce the quantum numbers of 
the SM quarks and leptons.  The value of this coupling at 
the energies of the $Z^0$ experiments is $\alpha_1 = 1/59.0$.

With this convention, the hypothesis of grand unification implies that the 
three couplings $\alpha_1$, $\alpha_2$, $\alpha_3$ have values at the 
mass scale of $\mz$ given in 
terms of a unification mass scale $M_U$ and a corresponding unification
coupling value $\alpha_U$ by the relation
\beq
   \alpha_i^{-1}(\mz) =   \alpha^{-1}_U -  {b_i\over 2 \pi} \log {\mz\over M_U}
                                        + \cdots\  .
\eeq{isrunning}
with
\beq
       b_1 = -{41\over 10}   \qquad b_2 = {19\over 6}   \qquad b_3 = 7
\eeq{SMbvals}
We can test this relation in two ways.  First, we can use 
\leqn{isrunning} and the
precisely known values of $\alpha_1$ and $\alpha_2$ to compute $\alpha_U$ and 
$M_U$, and then use these values to compute $\alpha_3$.  The result is
$\alpha_3 \approx 0.07$, in serious disagreement with \leqn{couplings}.
Second, we can eliminate  $\alpha_U$ and $M_U$ among the three relations
\leqn{isrunning}, to obtain the prediction
\beqa
    B = {b_3 - b_2 \over b_2 - b_1} &=& {\alpha_3^{-1} - \alpha_2^{-1}\over
        \alpha_2^{-1} - \alpha_1^{-1}}\CR
     &=& 0.717 \pm  0.008  \pm 0.03 \ ,
\eeqa{Bvalue}
where the first error is due to the experimental determination 
of the values of 
the $\alpha_i$ and the second is my estimate of the theoretical error from 
neglect of higher-order corrections in \leqn{isrunning} \cite{neglect}.  The 
coefficients \leqn{SMbvals} give  $B = 0.528$, again, in poor agreement 
with the
data.

\begin{figure}[tb]
\begin{center}
\includegraphics[width=.90\textwidth]{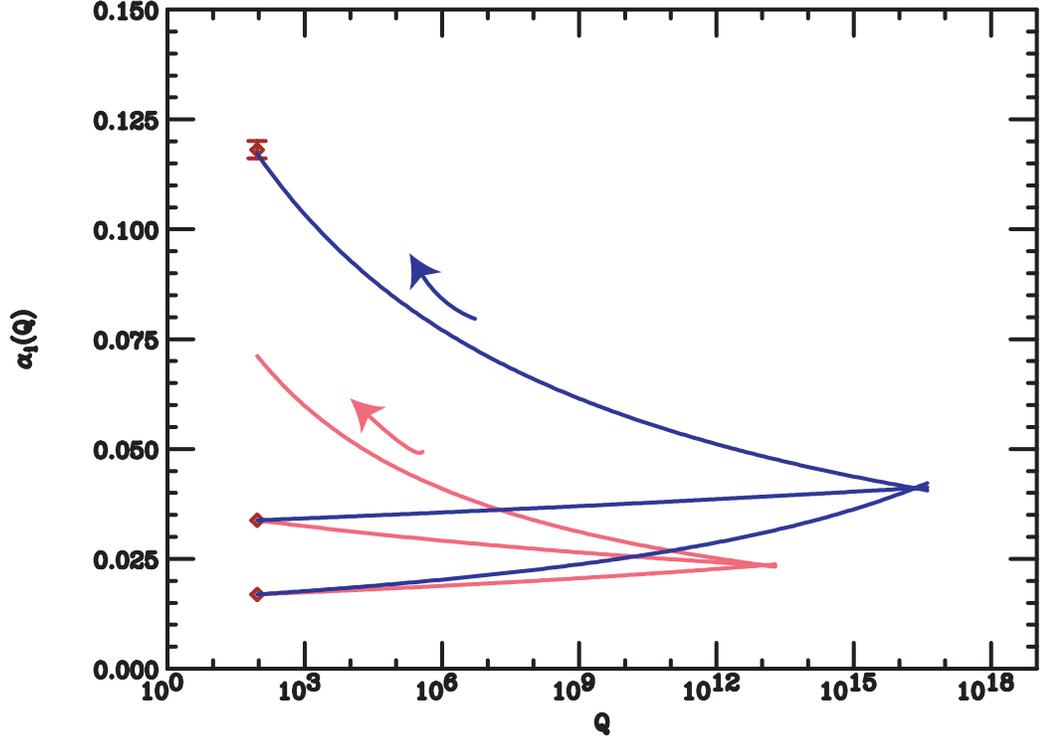}
\caption{Determination of $\alpha_3(\mz)$ from $\alpha_1(\mz)$ and 
     $\alpha_2(\mz)$ using the grand unification of couplings in the 
         SM and in its supersymmetric extension. The lower set of 
        three curves uses the $b_i$ values from the SM, the upper set
          those of its supersymmetric extension.}
\label{fig:GUTrelations}
\end{center}
\end{figure}

The determination of $\alpha_3$ from $\alpha_1$ and $\alpha_2$ is shown 
graphically as the lower set of curves in Fig.~\ref{fig:GUTrelations}. 
 A significant aspect of the calculation
is that the grand unification  scale turns out to be more than 10 orders of 
magnitude higher than the highest energy currently explored at accelerators.
If new particles appear at higher energy, their contributions will 
change the values
of the $b_i$.  If the SM is extended by the addition of supersymmetry, and if 
supersymmetry partners have masses within about an order of magnitude of
 $\mz$, the appropriate values of the $b_i$ to use in computing the predictions
of grand unification are those including the contributions from the 
supersymmetry
partners of quarks, leptons, gauge bosons, and Higgs bosons:
\beq
       b_1 =  -{33\over 5}  \qquad b_2 = -1   \qquad b_3 = 3 
\eeq{SUSYbvals}
These values give
\beq
          B = {5\over 7} = 0.714 \ ,
\eeq{BwithSUSYval}
in remarkable agreement with \leqn{Bvalue}.  The new evaluation of $\alpha_3$
is shown in Fig.~\ref{fig:GUTrelations}  as the upper set of curves.  The 
grand unification scale in this calculation is $M_U = 2 \times 10^{16}$ GeV, a 
value that is not so different (at least on a log scale) from the 
mass scale of 
quantum gravity.

The hypothesis of grand unification has implications for the properties of the 
Higgs boson.  Like the gauge couplings, the parameters that determine the 
mass of the Higgs boson vary as functions of the mass scale as the result of
quantum field theory corrections.  The effect of the corrections is always 
to lower
the prediction for the Higgs boson mass as the length of the extrapolation
 from 
the grand unification scale to the $Z$ scale is increased.  In a supersymmetric
grand unified theory with the value of $M_U$ just computed, it is difficult to 
arrange for a Higgs boson mass larger than 150 GeV.  Even extensive searches
have turned up  no such theory in which the Higgs boson mass is larger
208 GeV \cite{QEspin}.  This purely theoretical constraint on the Higgs boson 
mass corresponds nicely to the experimental constraint  \leqn{Hmassconstr}
discussed in Section 2.

\subsection{Dark matter and dark energy}

As I have already discussed, probes of the cosmological mass and energy
distribution indicate that the energy content of the universe is close to its
critical value $\Omega_0$.  About 30\% of this energy is composed of
nonrelativistic particles of non-baryonic matter.  About 70\% comes from 
the energy of the vacuum, or from some entity that behaves like vacuum energy
on the time scales of cosmological observations. 

Supersymmetry gives a natural candidate for the identity of the dark matter and
a mechanism for the survival of dark matter particles from the Big Bang.  
Consider the quantity
\beq
        R = (-1)^{3B - L + 2J} \ .
\eeq{defofR}
where $B$ is baryon number ($3B$ is quark number), $L$ is
 lepton number, and $J$ is spin.  This object
is constructed in such a way that all ordinary particles---leptons, baryons,
 mesons, gauge bosons, and even Higgs bosons---have $R = +1$.  The 
superpartners of these particles, however, have $R = -1$.  It is observed that
$B$ and $L$ are quite good symmetries, so it is not difficult to arrange 
that $R$
is conserved.  Then the lightest supersymmetry partner will be absolutely 
stable.  If this stable particle is the partner of the photon, or of the 
$U(1)$ 
gauge boson of $SU(2)\times U(1)$, it has all the properties required of a 
dark matter particle, being neutral, heavy, and weakly interacting.

The origin of the dark energy is more mysterious.  It is difficult in any 
current
theoretical framework to understand why the energy density of the vacuum is 
so small.  The spontaneous breaking of $SU(2)\times U(1)$ changes the 
energy density of the vacuum by an amount of order $\Delta \rho \sim \mh^4$.
However, the observed energy density is
\beq
         \rho_\Lambda \sim  (2 \times 10^{-14} \mh)^4 \ .
\eeq{therealrho}

Without supersymmetry, however, no one even knows how to begin.  In a 
non-supersymmetric theory, the energy of the vacuum is shifted by quantum
corrections in an arbitrary and uncontrolled way.  With supersymmetry, there 
is at least a natural zero of the energy.   It follows from 
\leqn{identifyP} that
\beq
             H = {1\over 4} \tr \left\{  Q_\alpha, Q^\dagger_\alpha \right\}  
\eeq{findH}
By \leqn{positivity}, the energy  is positive, and it is zero in a 
state $\ket{0}$
annihilated by $Q$ and $Q^\dagger$.  If supersymmetry is spontaneously 
broken, the vacuum energy becomes nonzero, but at least we know in 
principle where the zero is.

\subsection{Hints and anomalies}

At any given time, the data of elementary particle physics shows some small
deviations from the predictions of the SM that may or may not materialize
in the future into a real discrepancy.  I would like to highlight two current
anomalies that might be hints of the presence of supersymmetry. 

In the last few months of the operation of LEP, events accumulated that seemed
to be inconsistent with SM background and consistent with the production of 
a Higgs boson of mass about 115 GeV.  This was a marked contrast to 
previous experience at LEP, in which the observed event distributions had been
in excellent agreement with SM calculations.  However, the final
significance of the observation was only about 2 $\sigma$, statistically
unconvincing~\cite{KadoTully}. (Compare, for example, \cite{ALEPHHiggs}
and \cite{OPALHiggs}.)
I have already explained that supersymmetry typically implies a low mass 
for the
Higgs boson.  But this result is especially tantalizing because there 
is a stronger
upper bound on the Higgs boson mass in the `minimal' supersymmetric 
extension of the SM, the model with the minimum number of Higgs fields.  
In this
model, supersymmetry constrains the Higgs field potential in such a way 
that the
mass of the Higgs boson must be comparable to that of the $Z^0$. 
The Higgs boson mass must be less than 135 GeV, and for typical parameters
the value is between 90 and 120 GeV.

The Brookhaven Muon $g$-2  experiment has reported a discrepancy from the
SM of about 4 parts per billion~\cite{Muong}.
In a theory in which the supersymmetry partners of the leptons and the $W$ 
boson are both about 200 GeV,  this is roughly the expectation for the new 
contribution to the muon $g$-2 from radiative corrections containing 
these supersymmetric particles.  However, the status of this anomaly is 
still in question, because parts of the SM contribution to the 
muon $g$-2, the hadronic vacuum polarization and hadronic light-by-light
scattering diagrams, are not under control at the level of parts-per-billion
contributions~\cite{Nyff,Davier}.  As a result of this uncertainty, we 
can only say that the  significance of the anomaly is
somewhere between 1 and 3 $\sigma$. 

It will be interesting to see whether these anomalies are confirmed in the next
few years.

\section{Beyond the Supersymmetric Standard Model}

We have now seen that the addition of supersymmetry to the SM addresses
many of the major questions about that model that I have posed in Section 2.
For this reason, I consider it likely that supersymmetric partners of the SM
particle really do exist, and that they will be discovered at 
accelerators before
the end of the decade.   But this will only be the beginning of the path to the
next revolution in physics.  Let us now look at what lies further 
down this road.

I have already noted that, to describe Nature, supersymmetry must be a 
spontaneously broken symmetry.  Many aspects of the arguments given in 
the previous section that supersymmetry is relevant to particle physics depend
not only on the presence of the new symmetry but also on the values of the 
superpartner masses.  In the arguments given above, it is 
actually the scale of 
the supersymmetry-breaking mass parameters that determines the size of the 
Higgs mass and vacuum expectation value, and also the mass of the particles
of cosmological dark matter.

It is therefore important to investigate the mechanism of the spontaneous
breaking of supersymmetry.  The first place to look for this mechanism
is in the dynamics of the supersymmetric extension of the 
Standard Model.  However, this leads to a dead end.  Not only is there no 
obvious mechanism to be found, but there are good reasons why supersymmetry
breaking cannot come from physics directly connected to the Standard Model
particles.  For example, if an extension of the Standard Model contained a
tree-level potential that gave supersymmetry-breaking, the fermion and boson 
masses generated
by this model would obey the constraint 
\beq
      \tr  (m_f^2 - m_b^2) = 0
\eeq{mtrace}
This constraint would 
hold, not only for the whole spectrum, but also separately 
for each
charge sector.  Then, for example, there would need to be very light
squarks.
  More general constraints come from the strong bounds
on the supersymmetric contributions to quark mixing processes such as
the  $K^0$ or $B^0$ mixing amplitudes.  The superparticle  mass spectrum
must take a special form to avoid these contributions.  For example, it 
must be almost degenerate among squarks of the three generations.  It is 
not clear how dynamics in which the quark masses or other species-dependent
 couplings play
an important role can lead to such degeneracy.

Successful models of the supersymmetry spectrum start with a different 
strategy, assuming that supersymmetry breaking arises in a `hidden sector'
that is only weakly coupled to the Standard Model particles.  
The hidden sector is
assumed to couple through gauge bosons and gauginos, through supergravity,
or through other particles whose couplings can be sufficiently isolated
from the physics that leads to quark and lepton masses.

Where did this `hidden sector' come from?  What requires it?  Doesn't this 
constitute an unnecessary multiplication of hypotheses?

The answer to this question comes from string theory.  As I have 
discussed above, I do not insist that 
string theory is correct, but I am impressed that it does give an example of a 
theory that could, in 
principle, contain all of the interactions of Nature.  
  So it is worth taking
seriously what string theory has to say about the formulation of a `theory of 
everything'.

\begin{figure}[tb]
\begin{center}
\includegraphics[width=.95\textwidth]{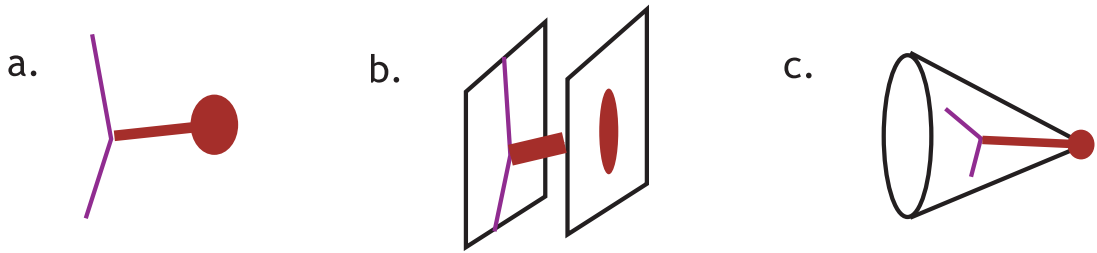}
\caption{Some pictures of the generation of masses for supersymmetric 
  particles by coupling to a `hidden sector' with spontaneous 
   supersymmetry breaking: a. from a gauge interaction outside the Standard
    Model; b. from a brane displaced into an extra space dimension; c. 
   from a sector of particles bound to a singularity in the compact manifold
       of extra dimensions.}
\label{fig:hidden}
\end{center}
\end{figure}

In fact, unified theories of Nature within string theory require a large 
superstructure.  String theory specifies the number of space-time dimensions
to be eleven.  The familiar four 
dimensions of space fill out part of this structure.
Part is taken up by curved space dimensions. These form compact 
manifolds  whose 
symmetries are the symmetries of the Standard Model gauge group and which,
by virtue of this, give rise to the Standard Model gauge bosons.  But there is
room for more.  Typical models of Nature built from string theory contain
additional gauge interactions from a variety of sources.  These can arise 
from additional symmetries of compactified extra dimensions.  They can also
arise in more subtle ways.  For example,  string theories contain as classical 
solutions hypersurfaces (called `branes')  with associated gauge bosons. 
Branes
can float freely in the extra dimensions or wrap around singularities or 
topological cycles of the compact manifolds that these directions form.  A
new non-Abelian gauge sector outside the Standard Model is potentially a 
source of new interactions that could break supersymmetry.  Since all 
parts of the model are linked by string interactions and  gravity, a new 
sector of this type would be a hidden sector in the sense of used earlier in 
this section.  In Fig.~\ref{fig:hidden}, I show some examples of hidden sectors
in extra dimensions whose weak coupling to the Standard Model fields can be 
understood geometrically.

The geometrical relations seen in Fig.~\ref{fig:hidden}
 determine the pattern of the
soft super\-sym\-me\-try-breaking parameters induced among the Standard Model
superpartners.  Some relatively simple schemes that generate simple but 
nontrivial patterns in the spectrum are described in 
\cite{Horava,RandallSundrum,SchmaltzBr}.   More complicated---and perhaps 
more realistic---patterns due to the geometry of supersymmetry breaking
remain to be discovered.  Conversely,
the evidence of this geometry, or of some more
subtle picture of supersymmetry breaking, is present 
in the patterns that can be
observed in the superpartner mass spectrum.  These traces of physics at
extremely small distances are waiting there for us to tease them out.

\section{Interpretation of the SUSY-breaking parameters}

In the previous two sections, I have argued that supersymmetric particles must
be light---light enough to be discovered at the next generation of particle 
accelerators.  I have also argued that their mass spectrum will be 
interesting to
study, because its regularities encode information about the geometry of space
at very short distances.  However, there is a complication in obtaining this 
information that should be discussed.  The observed masses do not 
fall simply into the pattern of the underlying SUSY-breaking parameters.  
Rather, they 
are modified by quantum field theory effects that we must disentangle.

In Section 4.2, I explained that the Standard Model coupling constants, which
appear to be unequal by large factors, actually have the same value at the 
scale of grand unification.  The couplings are then modified by different 
amounts when we analyze their influence on measurements at length scales
much larger than the grand unification scale.  After measuring these couplings
with precision, however, we can perform the analysis shown in 
Fig.~\ref{fig:GUTrelations} and discover the regularity.  The 
supersymmetry-breaking
mass parameters have a similar difficulty.  They are changed substantially
from the enormous energy scale where they are created to the much 
lower energy scale of accelerator experiments where they can be observed.
Fortunately, the changes are predicted by quantum field theory, 
so it is possible
here also to undo their effect by calculation.

The gauginos, the superpartners of the gauge bosons, 
obey a simple scaling relation.  To leading order, they
 are rescaled by the same factor as the Standard Model 
gauge couplings.  So if, for example, the masses $m_1$, $m_2$, and 
$m_3$ of the $U(1)$, $SU(2)$, and $SU(3)$ gauginos are equal to a common 
value $m$  at the energy scale $M$ of 
grand unification, then at any lower energy scale $Q$ these 
parameters will obey 
the relation
\beq
      m_i(Q) =   {\alpha_i(Q)\over \alpha_i(M)} m \ .
\eeq{mirunning}
This simple consideration predicts that the three mass values have the ratio
\beq
    m_1 \ : \ m_2 \ : \ m_3 \ =   0.5 \ : \ 1 \ : \ 3.5  
\eeq{mirat}
for the physical values at accelerator energies.
  The corresponding relation for the
supersymmetry partners of quarks and leptons is more complicated.  Quantum
field theory predicts an additive contribution resulting 
from the fluctuation of a squark
or slepton into the  corresponding quark or lepton plus a
 massive gaugino.  The 
squarks couple relatively strongly to the gluino, and that
particle is also expected to receive a larger mass from 
\leqn{mirat}, so this mechanism typically makes the squarks
 heavier than the sleptons.  In the extreme case in which the 
squarks and sleptons have zero mass at the grand unification scale, 
the physical masses at the 
TeV scale should be in the ratio
\begin{eqnarray}
& &    m(\s e_R) : m(\s e_L) : m(\s d_R) : m(\s u_R) : m(\s u_L/\s d_L) : 
          m_2 \CR
 & & \hskip 0.3in 
   =      0.5 : 0.9 :  3.09 :  3.10 :  3.24:  1 \   .
\label{sqarkms}\end{eqnarray}

\begin{figure}[tb]
\begin{center}
\includegraphics[width=.60\textwidth]{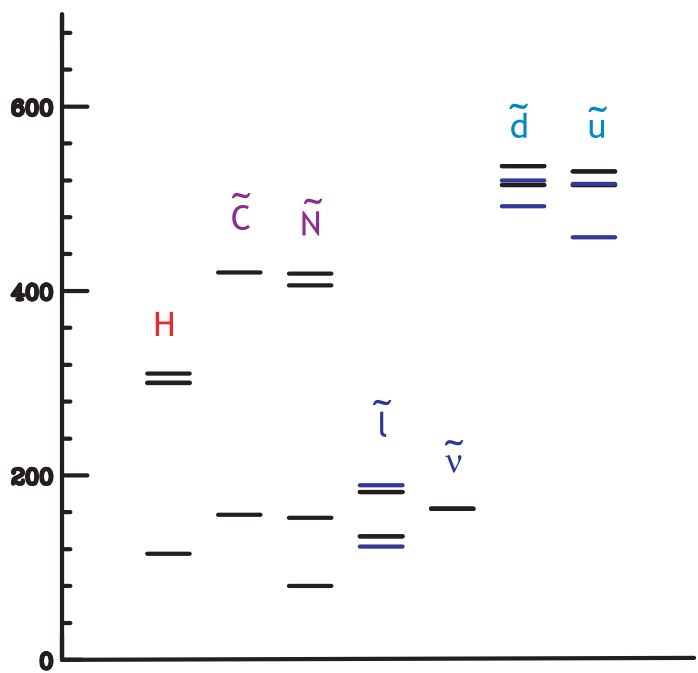}
\caption{Sample spectrum of supersymmetric partners, based on universal 
 masses for gauginos and sfermions at the energy scale of grand unification.}
\label{fig:spectrum}
\end{center}
\end{figure}
\begin{figure}[p]
\begin{center}
\includegraphics[width=.95\textwidth]{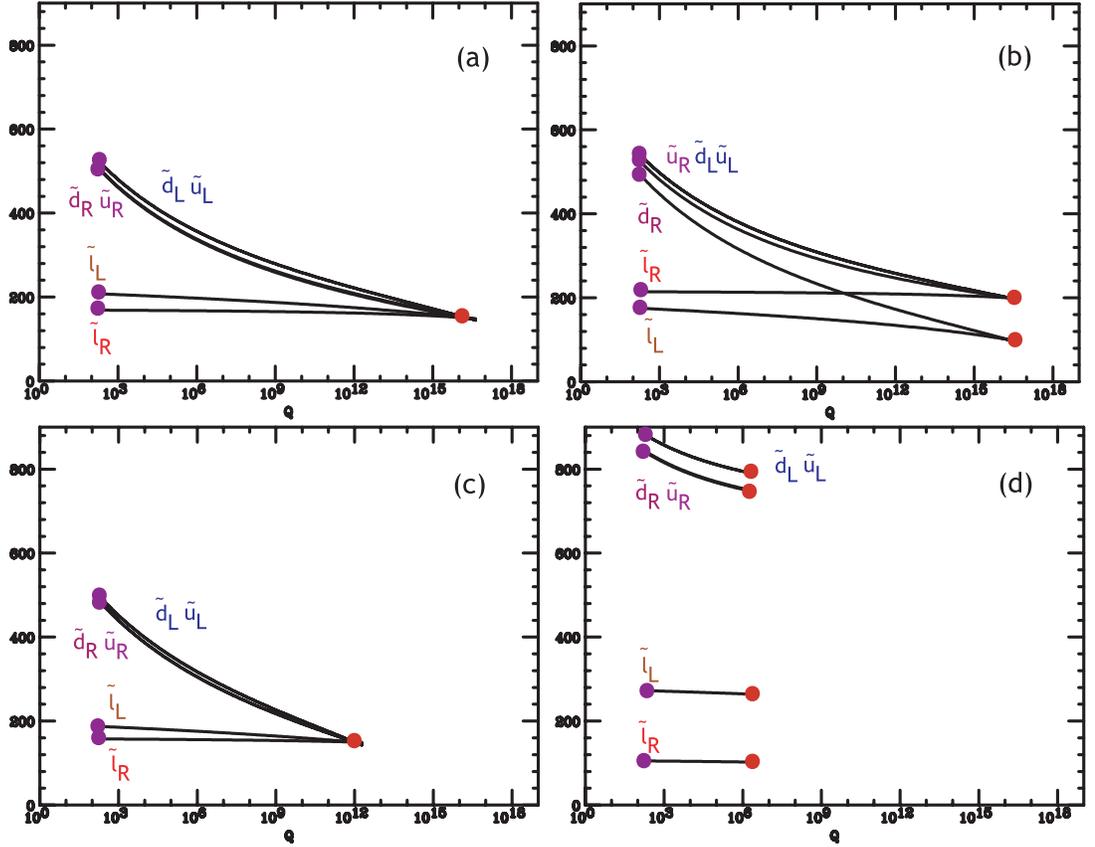}
\caption{Evolution of squark and slepton masses from the mediation scale
 $M$ down to the weak interaction scale (100 GeV) in four different 
scenarios:  (a) universal mass at $M$ equal to the grand unification 
scale; (b) separate masses for each individual $SU(5)$ mutliplet at $M$; 
(c) universal mass at $M$ well below 
the grand unification scale;  (d) masses generated at a low mediation 
scale $M$ by Standard Model gauge and gaugino couplings.  The dots on the
right are the underlying parameter values; the dots on the left are the 
masses that would be measured in experiments.}
\label{fig:evolution}
\end{center}
\end{figure}

A complete spectrum for the superparticles that illustrates these 
features is shown in Fig.~\ref{fig:spectrum}. In this spectrum, 
I have assumed a common mass for the gauginos and a separate common
 mass for the squarks and sleptons.  The mass splittings between the
 squarks and sleptons and between the electroweak and strong-interaction 
gauginos come from quantum field theory corrections.  
This assumption is the simplest one possible---and, probably, much too simple.
In Fig.~\ref{fig:evolution}, I illustrate some alternative hypotheses 
for the underlying supersymmetry-breaking parameters.   The figures
show the quantum field theory evolution of parameters from the original
supersymmetry-breaking parameters on the right to the measurable values of
squark and slepton masses on the left.
It is a common 
feature that the squarks are heavier and somewhat degenerate, while the
slepton partners of right- and left-handed leptons are lighter
and well split in mass.   Precision analysis of the 
spectrum is needed to go beyond this qualitative feature, but the figure 
indicates that the detailed predictions for the supersymmetry spectrum
do vary significantly in a way that 
can reveal the 
differences in the original assumptions.

Some other properties of 
 the spectrum should also be noted.  The partners of the 
heaviest quarks and leptons $\tau$, $b$, and $t$ are split off from the 
others by two effects.  First, there is an additional quantum field theory
contribution  due to
the coupings to the Higgs bosons that are responsible for the larger masses of
the quarks and leptons.  Second, there are supersymmetry-breaking contributions
to the sfermion-sfermion-Higgs couplings that lead to mixing between
 the partners of the left- and right-handed fermion species. 

Mixing of particle states is an  issue in many parts of the 
supersymmetry spectrum, and one that significantly complicates 
the interpretation of the particle masses.     
Not only do the two scalar partners of each heavy quark or lepton 
mix together, but also there can be important mixings among the 
partners of the gauge bosons and Higgs bosons.  In addition to the 
$W^+$ partner $\s w^+$, there is a fermionic 
partner of the Higgs boson $h^+$; after electroweak symmetry breaking, 
these particles have the same quantum numbers and can mix.  
The mass eigenstates of this system, which are the observable physical
particles, are called `charginos', $\s C^+_i$; they are quantum-mechanical
mixtures of the two original states.  Typically, 
one mass 
eigenvalue is  close to $m_2$  while the other is close to an underlying
 Higgs mass parameter $\mu$.  To determine either parameter with precision, 
the mixing must be understood.  Similarly, the gaugino partners of the 
photon and the $Z^0$ combine
with two neutral Higgs fermions to form a four-state mixing problem that must
be disentangled.  The mass eigenstates of this mixing problem
are called `neutralinos', $\s N^0_i$.

In addition to their role in the precision analysis of spectra, the mixing
 parameters just described are of interest in their own right.  To check 
the story I have told in 
Section 4.1 about the origin of electroweak symmetry breaking,  we should 
use the
measured values of the supersymmetry parameters to compute the Higgs
 boson vacuum
expectation value.  The parameters of $\s t$ mixing turn out to play an 
important role in this calculation, as does the parameter $\mu$.  The 
mixing parameters also play an important role in the calculation of the 
abundance of cosmological dark matter left over from the early universe. 
 In Section 4.3, I have identified the dark matter particle with the lightest
 neutralino, $\s N^0_1$. The reaction cross sections of this particle 
depend on the composition of the lowest mass eigenstate of the four-state 
mixing problem of neutral fermions.   In addition, the pair 
annihilation of neutralinos often is dominated by the annihilation 
to tau lepton pairs, which brings in the mixing problem of the tau 
lepton partners.  Both sets of mixing angles need to be measured 
before we can produce a precise prediction for the dark matter density
from supersymmetry that we can compare to the measured cosmological 
abundance.

\section{Measuring the Superspectrum}

The complications discussed in the previous section add some 
difficulty to the interpretation of the supersymmetry spectrum, 
but these difficulties are no worse
than those typically encountered in atomic or nuclear spectroscopy. They are a 
hint that the experimental determination of the underlying parameters of 
supersymmetry will be a subtle and fascinating study.  

A serious question remains,
though, about whether we can actually have the data.  The properties of 
supersymmetric particles cannot be determined on a lab bench.  
High energies are required, and also a setting in which the properties of
the exotic particles that are produced can be well measured.  Cosmic
rays could potentially provide the required energies, but they do not 
provide enough rate.  To produce massive particles, the quarks or gluons
inside colliding protons must come very close together, and this means that
the typical cross sections for producing supersymmetric particles in 
proton-proton collisions are less than $10^{-10}$ of the proton-proton
total cross section.  The only known technique for extracting enough of these
rare events from very high energy collisions is that of creating controlled 
reactions at dedicated particle accelerators.

Though it might be possible to glimpse supersymmetry at the currently
operating accelerator at Fermilab,  a comprehensive study
of supersymmetry spectroscopy will require new accelerators with both higher
energy and greater capabilities than those that are now operating.  The 
high energy physics community is now planning for these
 accelerators---the Large Hadron Collider (LHC) at CERN and  
a next-generation electron-positron collider along the lines 
of the TESLA project in Germany or the 
NLC and JLC projects in the US and Japan.  In this section, I will 
review 
some of the experiments at these facilities that might follow the discovery 
of supersymmetric particles.  

Even given the needed energy and rates of 
particle production, it is a nontrivial question whether accelerator
experiments can be sufficiently incisive to allow us to work out the 
detailed properties of the supersymmetry spectrum.  But, in the next 
several sections, I will argue that it is so.
Despite the fact that 
 experiments  at these proposed facilities are far removed from the human 
scale, 
they can include many subtle analytic methods.
We can have the data 
to recover and understand the basic parameters of supersymmetry.   It 
will be an 
adventure to perform these experiments and lay out the spectroscopy of 
supersymmetric particles---and another adventure to interpret this spectrum in 
terms of the physics or geometry of deep underlying distance scales.

\subsection{Experiments at the LHC}

The LHC is a proton-proton collider, with a center-of-mass energy of 
14~TeV, now under construction at CERN.  At energies so far above the proton 
mass, proton-proton collisions must be thought of as collisions of the 
proton's constituents, 
quarks and gluons.  The dominant processes are those from gluon-gluon 
collisions. 
Such collisions bring no conserved quantum numbers into the reaction 
except for the basic `color' quantum numbers of the strong interactions. 
 Thus, they can 
produce any species of strongly-interacting particle, together 
with its antiparticle,
up to the maximum mass allowed by energy conservation.

In the sample spectra shown in Fig.~\ref{fig:evolution}, the 
strongly-interacting 
supersymmetric partners,  the squarks and gluinos, are the heaviest 
particles in 
the theory.  These particles are unstable, decaying to quarks and to
the partners 
of the electroweak gauge bosons.  Often, the decays of the heavy particles 
proceed in several stages, in a cascade.  If the quantum number $R$ presented 
in Section 4.3 is conserved, the lightest supersymmetric partner 
produced in each 
cascade decay wil be stable and will exit the detector unobserved, 
carrying away 
some energy and momentum from the reaction.  These are the particles of
cosmological dark matter, and in the laboratory too they appear only 
as missing mass and energy.

These properties give the LHC events which produce supersymmetric particles a 
characteristic form.  Typical proton-proton collisions at the LHC are glancing 
collisions between quarks and gluons.  These produce a large number of 
particles, but these particles are mainly set moving along the 
direction of the proton beams, 
with relatively small perpendicular (or `transverse') momentum.  When heavy 
particles are 
produced, however, the decay products of those particles are given transverse 
momenta of the size of the particle mass.  A quark produced with large
transverse momentum materializes in the experiment as a cluster of 
mesons whose momenta sum to the momentum of the original quark and whose
directions are within a few degrees of the original quark direction.  
Such a cluster,
called a `jet', is the basic object of analysis in experiments at proton 
colliders.
Events with supersymmetric particle production contain multiple jets with 
large transverse momentum, and also unbalanced or missing transverse
momentum carried away by the unobserved stable dark matter particles.

\begin{figure}[tb]
\begin{center}
\includegraphics[width=.65\textwidth]{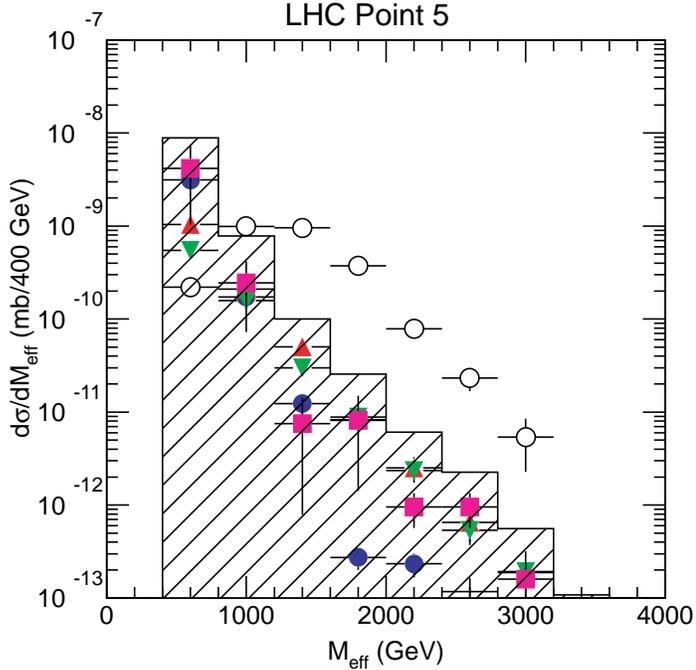}
\caption{Expected distribution of the quantity $M_{eff}$, 
defined by \leqn{Meffdef}, 
     in the ATLAS experiment at the LHC, from 
    Standard Model events and from events with supersymmetric particle 
      production, from \cite{ATLAS}.}
\label{fig:ATLASMeff}
\end{center}
\end{figure}
\begin{figure}[tb]
\begin{center}
\includegraphics[width=.65\textwidth]{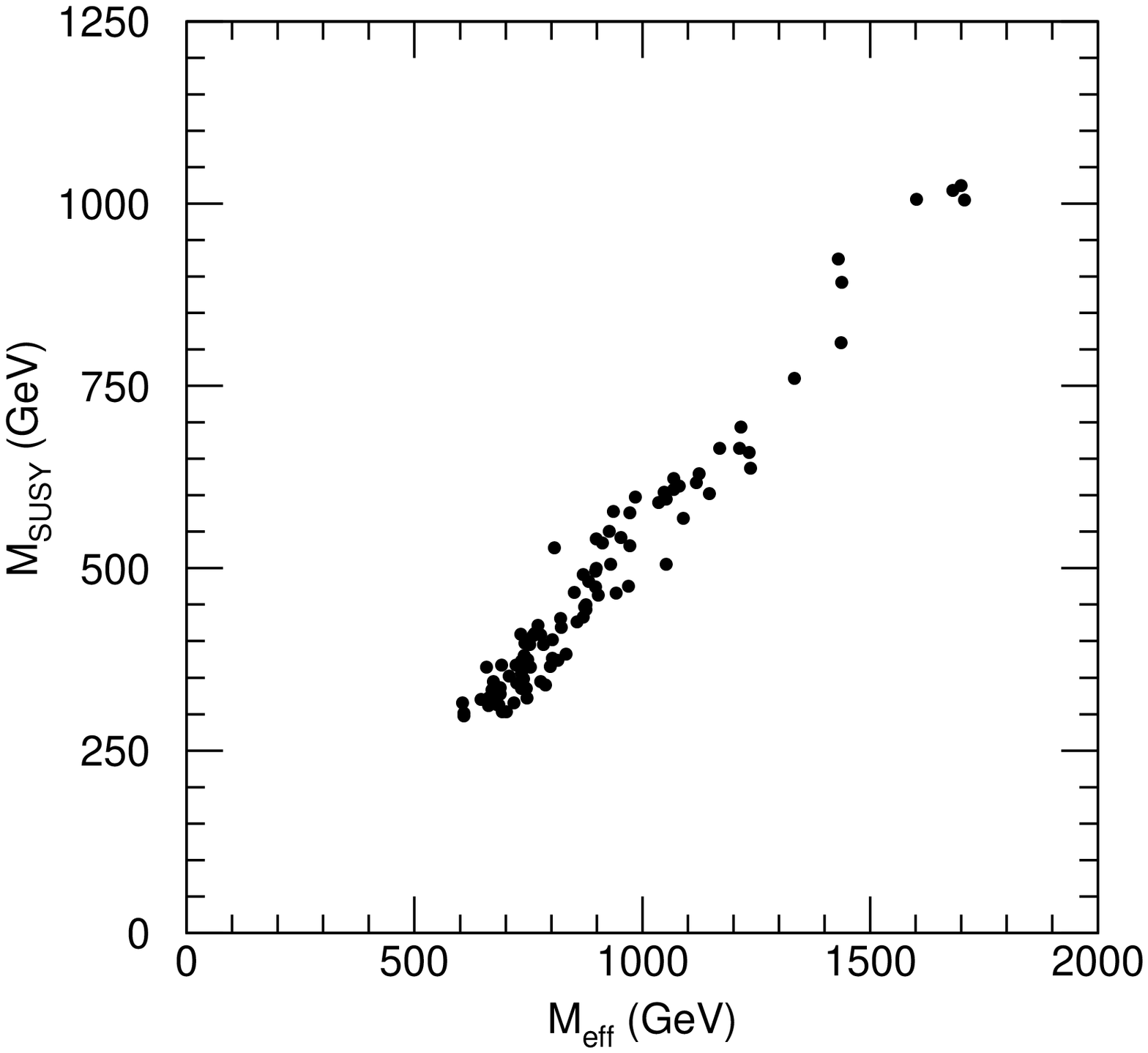}
\caption{Correlation of $M_{eff}$ with the lighter of the squark and 
 gluino masses,  from \cite{ATLAS}.}
\label{fig:Meffcorr}
\end{center}
\end{figure}
Studies of supersymmetry production carried out by the ATLAS experiment
at the LHC make use of a variable that is sensitive to all of these effects.  
Define
\beq
 M_{eff} =   \not p_T  + \sum_1^4  p_{Ti}  \  ,
\eeq{Meffdef}
the scalar sum of the $p_T$ imbalance and the $p_T$ values of the four 
observed jets of largest $p_T$.  Events with large $M_{eff}$  
come from new physics 
processes outside the Standard Model.  This is shown in 
Fig.~\ref{fig:ATLASMeff},
in which the $M_{eff}$ distribution expected from Standard Model events is
compared to that expected from supersymmetry production for one specific 
choice of the spectrum.  Not only can one use the variable $M_{eff}$ to select
events with supersymmetry, but also the average value of $M_{eff}$ is well
correlated with the mass of the strongly-interaction supersymmetric particles. 
This is shown in Fig.~\ref{fig:Meffcorr}, which gives a scatter plot 
of the average 
value of $M_{eff}$ versus the lighter or the squark and gluon masses for 
a number
of  supersymmetry spectra considered in the ATLAS study.

Once the mass scale of the supersymmetry spectrum is known and a sample of 
events can be selected, the more detailed properties of these events can give 
precise measurements of some of the spectral parameters.  The observables that
are most straightforward to measure are the energy and momenta of jets and 
leptons produced in the event, and these often do not have an unambiguous
interpretation.  However, in some cases, these parameters tell a very specific 
story.  Consider, for example, a spectrum in which the mass difference 
between the second and the lightest neutralino is less than the mass of 
the $Z^0$ boson.
Then the $\s N_2^0$ can decay to the light unobserved particle $\s N_1^0$ 
by
\beq
          \s N_2^0 \to \s N_1^0 + \ell^+ \ell^- \ ,
\eeq{Ntwodecay}
where $\ell$ is a muon or an electron.  Because there is not enough 
energy from the mass difference to form a $Z^0$, the system of two 
leptons has a broad 
distribution in mass.  However, it cuts off sharply at the kinematic endpoint
\beq
     m(\ell^+\ell^-) =   m(\s N_2^0 ) - m( \s N_1^0 ) \ . 
\eeq{mdiff}
By identifying this feature, it should be possible, in a scenario of 
this type, to 
measure the mass difference of neutralinos to better than 1\%.  The decay of
$\s N_2^0$ to $\s N_1^0$ is a typical transition at the last stage of the decay
cascade of the partners of left-handed quarks.

\begin{figure}[tb]
\begin{center}
\includegraphics[width=.95\textwidth]{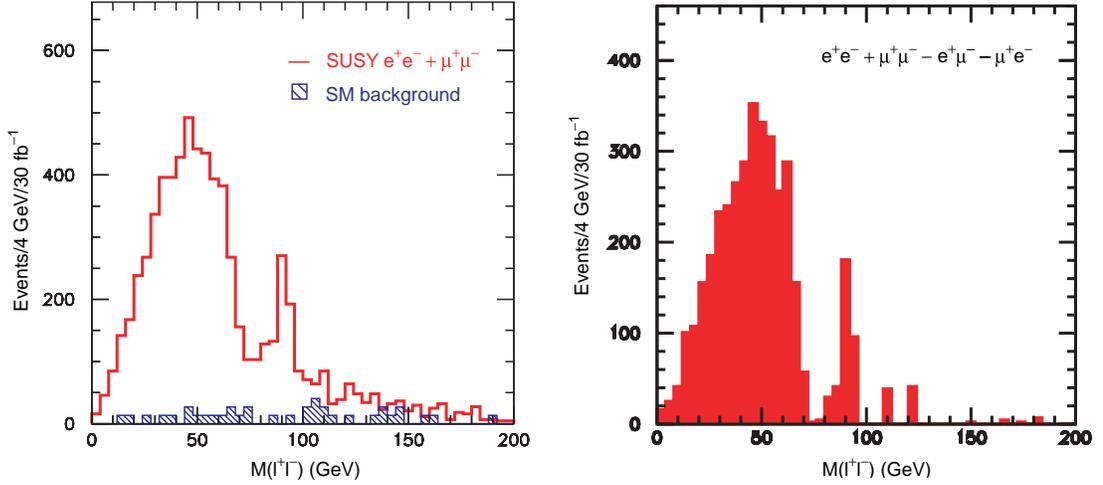}
\caption{Expected mass spectrum of $\ell^+\ell^-$ pairs in the ATLAS 
      experiment at the LHC, for the supersymmetry point 4 considered
      in \cite{ATLAS}.}
\label{fig:ellell}
\end{center}
\end{figure}
\begin{figure}[tb]
\begin{center}
\includegraphics[width=.5\textwidth]{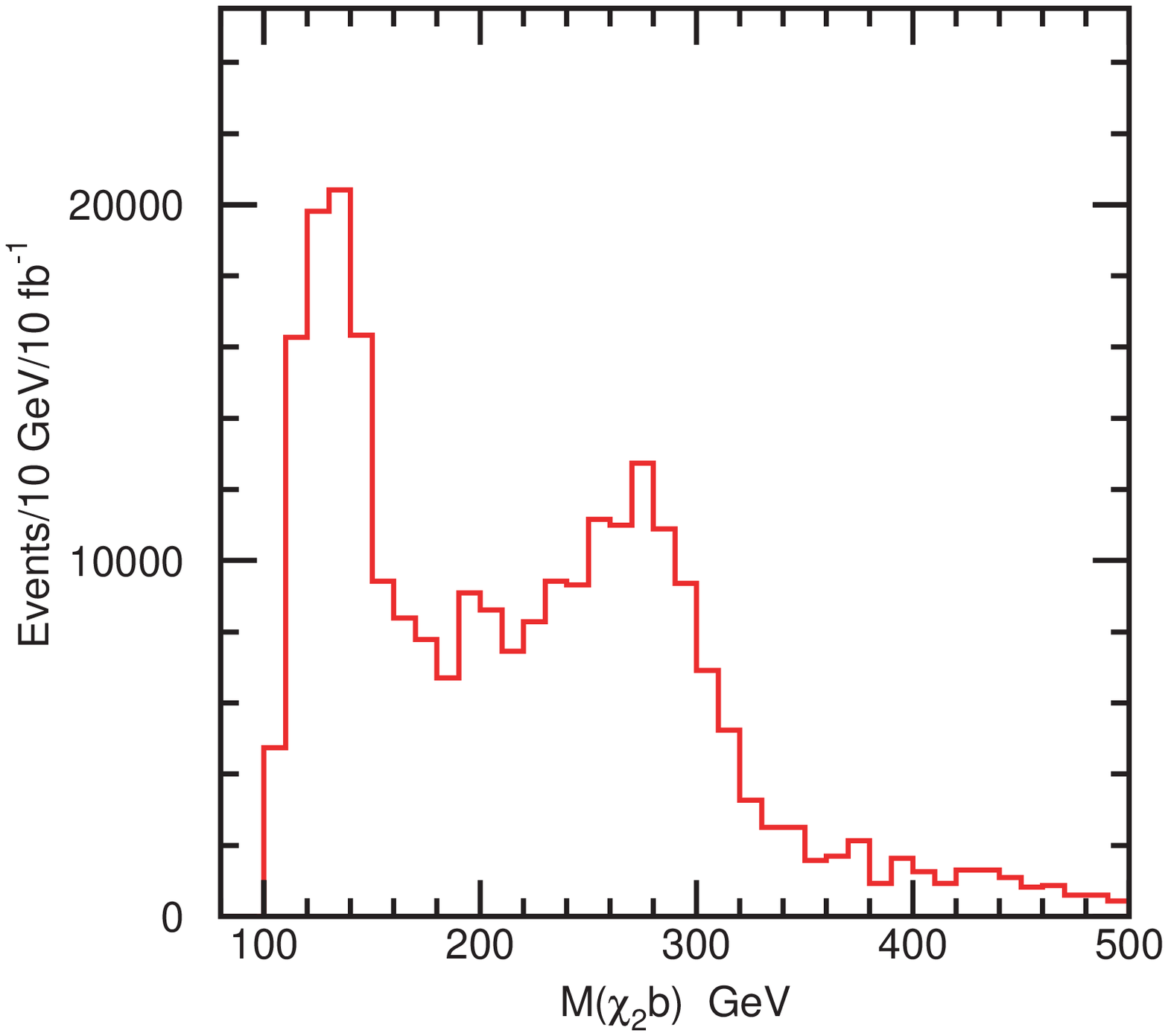}
\caption{Reconstruction of the $\s b$ mass by combining a reconstructed
       $\s N_2^0$ with a $b$ quark jet, from  the simulation study of 
          point 4 in   \cite{ATLAS}.}
\label{fig:bsquark}
\end{center}
\end{figure}
The $\ell^+\ell^-$ endpoint determination is illustrated in 
Fig.~\ref{fig:ellell}, which gives the lepton pair spectrum at 
one of the points studied by ATLAS.  The background from Standard 
Model processes is shown explicitly in Fig.~\ref{fig:ellell}(a); 
there is very little.  The observed leptons in the selected event 
then arise dominantly from supersymmetry decays, but from a number
 of different mechanisms.  Most of these mechanisms, however, produce 
charged leptons singly (with neutrinos) and
therefore produce one electron and one muon as often as a pair.  By subtracting
\beq
     (e^+e^-) + (\mu^+\mu^-)  - (e^+\mu^-) - (\mu^- e^+)
\eeq{substr}
we can concentrate our attention on the leptons produced in pairs.  The
 subtracted
mass spectrum is shown in Fig.~\ref{fig:ellell}.  The pairs with  mass of 
about 90 
GeV arise from decays of the third and fourth neutralinos by 
emission of a $Z^0$ boson, which then decays to $\ell^+\ell^-$.  
The peak
at lower mass comes from the  $\s N_2^0$ decays.  The endpoint is very sharp,
allowing a precise mass difference to be determined.

In many cases, this step is just the beginning of a deeper investigation. 
The events
near the endpoint in the mass distribution correspond to the  special 
kinematics in which the final $\s N_1^0$ is almost at rest in the 
frame of the $\s N_2^0$.  This
allows the maximum amount of the energy of the $\s N_2^0$ to go into the 
leptons, creating the maximum mass.  But this means that, if we can 
determine the mass of the $\s N_1^0$ from another set of measurements, 
we have the entire momentum vector 
of the $\s N_1^0$, and therefore the momentum vector of the $\s N_2^0$.  
If the 
$\s N_2^0$ was produced in a decay $\s q \to q \s N_2^0$, we can add the 
momentum of an observed quark jet and attempt to reconstruct the mass of the 
parent squark.  Fig.~\ref{fig:bsquark} shows an example of such an analysis.
The mass peak at about 270 GeV is the reconstructed squark; its mass is 
determined in this analysis to percent-level accuracy.

\begin{figure}[tb]
\begin{center}
\includegraphics[width=.7\textwidth]{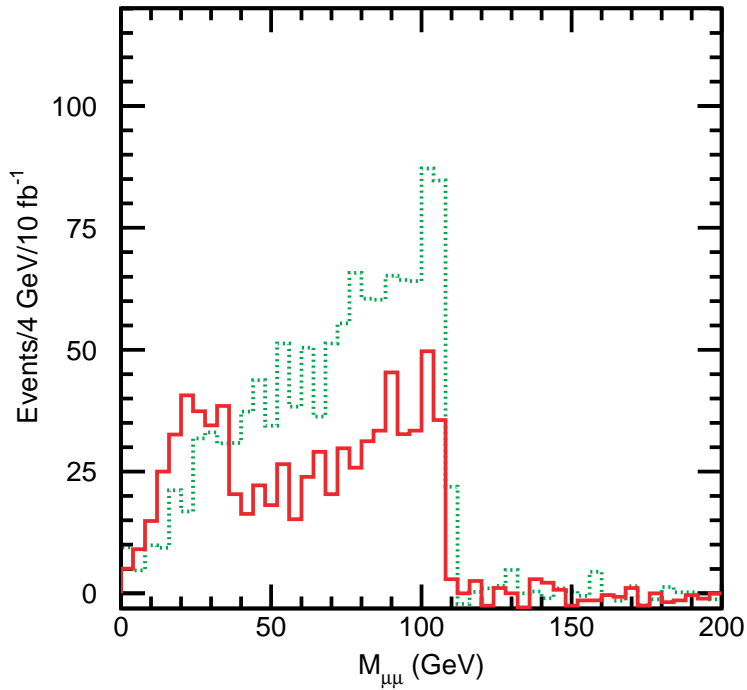}
\caption{Expected mass spectrum of $\ell^+\ell^-$ pairs in the ATLAS 
      experiment at the LHC, for a  supersymmetry parameter set in which 
      $\s N_2^0$ can decay to both $\s\mu$ states, compared to the 
       mass spectrum (shaded) in which only the decay to the lighter
        $\s\mu$ is allowed, from  \cite{ATLAS}.}
\label{fig:twoells}
\end{center}
\end{figure}
Less straightforward possibilities can also occur.  
Figure~\ref{fig:twoells} shows the $\ell^+\ell^-$ mass
 spectrum at another point 
considered in the ATLAS study in which 
the $\s N_2^0$ decays to $\s \ell \ell$.  It might happen that the 
$\s N^0_2$ has a kinematically allowed decay only to $\s\ell_R^\pm \ell^\mp$.
In other scenarios, the $\s N^0_2$ could decay to either the $\s\ell_L$ or
the $\s\ell_R$.  The latter case is shown as the solid curve in 
Fig.~\ref{fig:twoells}, with two sharp endpoints visible.
  There is obviously 
some subtlety in determining the correct decay pattern of the 
neutralinos from the data.  But the 
clues are there, and, if they are deciphered correctly, many parameters of the 
supersymmetry spectrum can be obtained.  More examples are given in 
ref.~\cite{ATLAS}.

\subsection{Experiments at the Linear Collider}

Experiments in electron-positron annihilation should present a quite different
view of the supersymmetry spectrum.   Electrons and positrons are elementary
particles, so they can annihilate to a state of pure energy without 
leaving over any
residue.  This state, like that produced by a gluon-gluon collision, 
is completely
neutral in its quantum numbers.   So an electron-positron collision 
can directly 
produce particle anti-particle pairs of any particle with
 electromagnetic or weak
interaction quantum numbers:  
\beq
      \ee \to X \bar X \ .
\eeq{basicxxbar}
The particles are produced back-to-back, each with the original
 electron energy.
It is even possible to control the spin orientations of the particles:
  In a linear
accelerator, the electron can be given a definite longitudinal 
polarization which is preserved during
the acceleration process.  Then the $X\bar X$ system  is produced 
in annihilation 
with angular momentum $J=1$, oriented parallel to the electron 
spin direction.

Because electrons and positrons radiate more copiously than protons, 
it is more 
difficult to accelerate them to very high energy.  So the energies 
planned for the 
next-generation electron-positron collider are much lower than that 
of the LHC, 
500 GeV in the first stage, increasing with upgrades to about 1 TeV.  
 This should
be enough energy to produce the lightest states of the superspectrum 
and subject
them to a controlled examination.

\begin{figure}[p]
\begin{center}
\includegraphics[width=.8\textwidth]{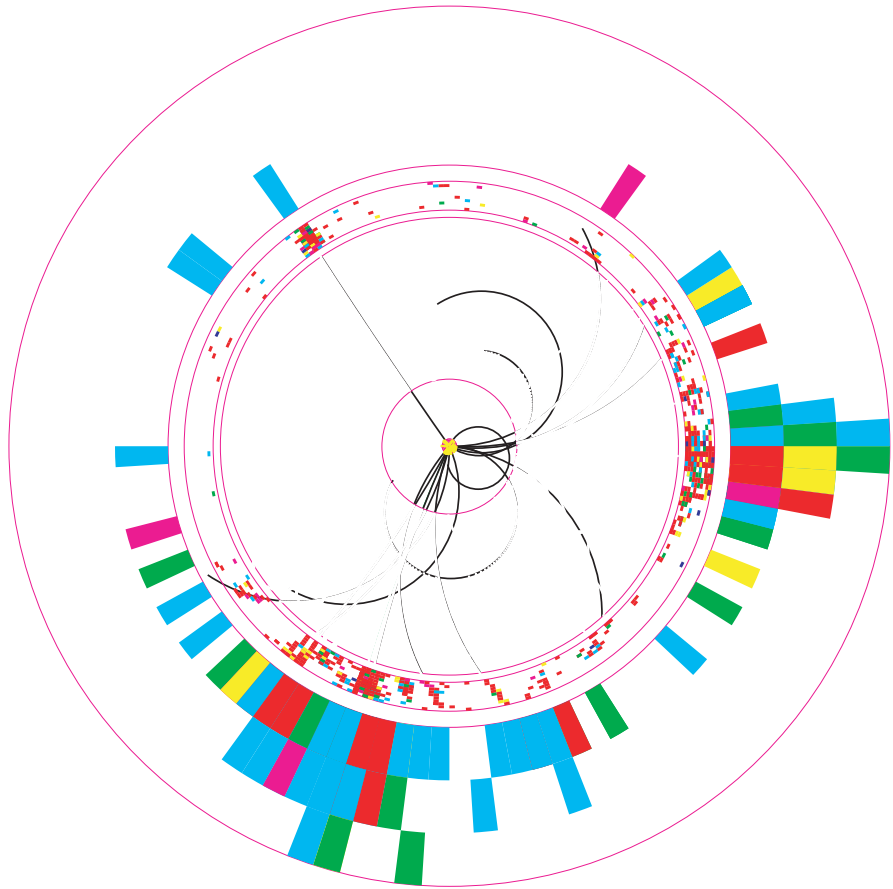}
\caption{A simulated event of $\ee$ annihilation to a chargino pair, 
as it would
       appear in a detector at  a linear $\ee$ collider, from \cite{Graf}.}
\label{fig:myCCbar}
\end{center}
\end{figure}
An example of a simulated supersymmetry event at this facility is shown in 
Fig.~\ref{fig:myCCbar}.  The reaction shown is the production of a pair of 
charginos, which subsequently decay to the lightest neutralino plus a pair of 
quarks or leptons:
\beq
   \ee \to \s C^+  \s C^-  \to  e^+ \nu \s N_1^0 \quad q \bar q \s N^0_1 \ .
\eeq{myCCbar}
The electron is visible as the isolated stiff  track.  
There are two well-defined jets which are the signals of the 
quark and antiquark.
The colored cells denote the 
energy deposition by both charged and neutral particles.   
The momentum and 
energy flow from the electron and the jets is simple and 
readily reconstructed, 
giving a clear picture of the whole event.

\begin{figure}[tb]
\begin{center}
\includegraphics[width=.5\textwidth]{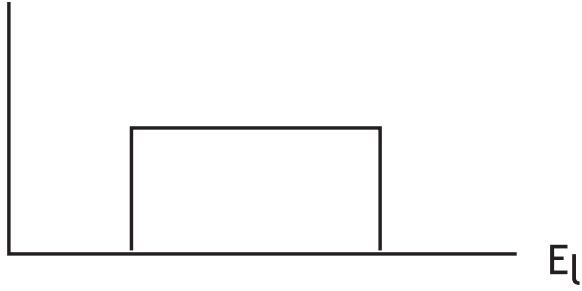}
\caption{Schematic form of the lepton energy distribution in slepton
  pair-production events.}
\label{fig:sleptonenergies}
\end{center}
\end{figure}
\begin{figure}[tb]
\begin{center}
\includegraphics[width=.95\textwidth]{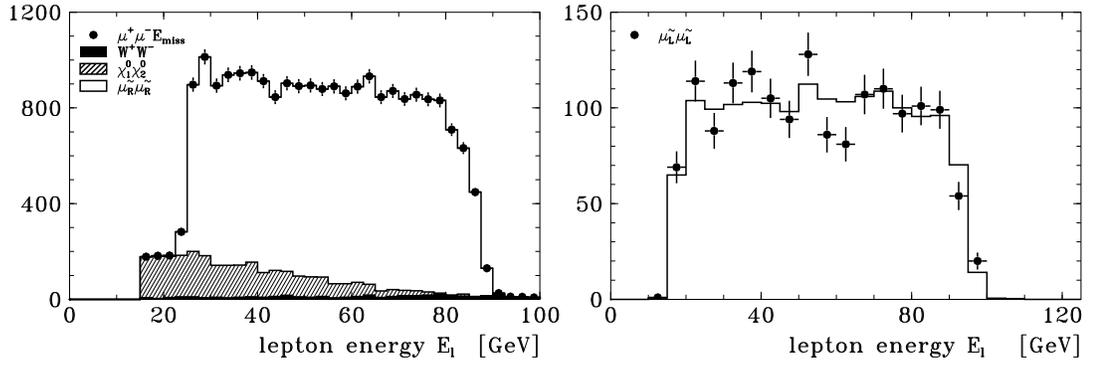}
\caption{Distributions of $\mu^\pm$ energy from simulations of 
               smuon decays in 
     smuon pair-production events, from \cite{Blair}.}
\label{fig:sleptons}
\end{center}
\end{figure}

The relation between the momenta of the decay products and the momenta of
the parent supersymmetric particles is also very simple.   
The cleanest correspondence comes in the case of slepton
 pair production.  The slepton decays to the corresponding 
lepton and a neutralino, for example,
\beq
   \s \mu \to \mu \s N_1^0\ .
\eeq{smudecay}
Because the slepton has spin 0, the decay is isotropic in its rest frame.  The 
sleptons are produced in motion, but the 
boost of an isotropic distribution is a distribution that is 
constant in energy between
he kinematic endpoints.  So the distribution observed in the lab has the 
schematic form shown in Fig.~\ref{fig:sleptonenergies}.    From the values of
the energy at the two endpoints, one can solve algebraically for the mass of
the slepton and the mass of the neutrinalino produced in the 
decay~\cite{Tsukamoto}. The masses can be determined by this technique 
to better than~1\%. 

\begin{figure}[tb]
\begin{center}
\includegraphics[width=.5\textwidth]{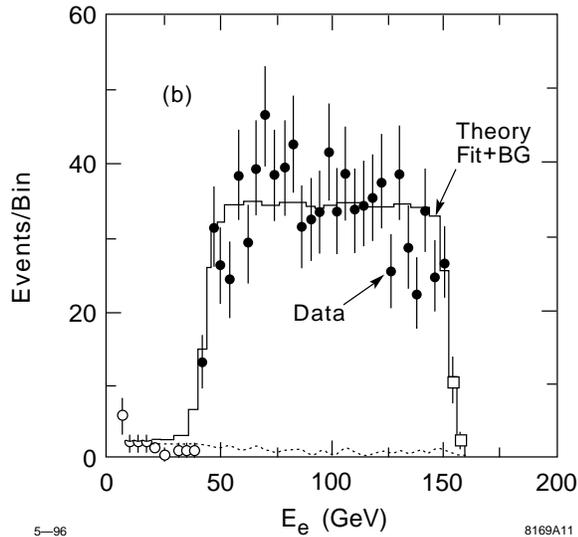}
\caption{Distribution of $e^-$  energy from $\s \nu$ decays in 
         a simulation of sneutrino pair-production events, from~\cite{Baer}.}
\label{fig:snu}
\end{center}
\end{figure}

\begin{figure}[tb]
\begin{center}
\includegraphics[width=.90\textwidth]{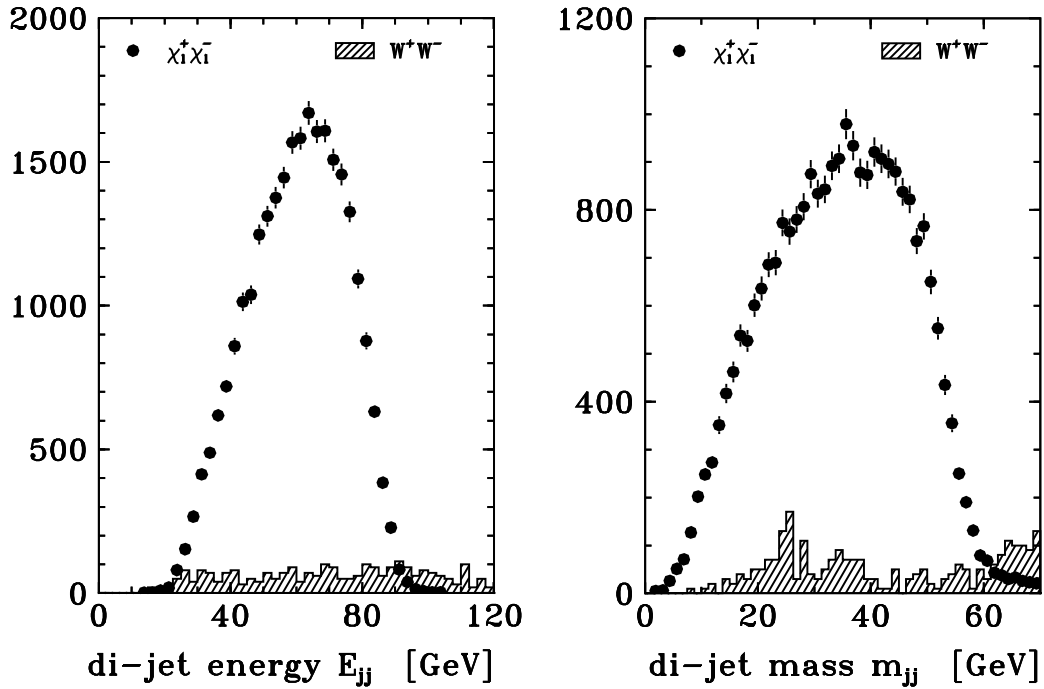}
\caption{Distributions of $q\bar q$ energy and mass distributions
       in a simulation of  chargino pair-production events,  
                     from \cite{Blair}.}
\label{fig:Cqqbar}
\end{center}
\end{figure}

In Fig.~\ref{fig:sleptons}, I show the energy distributions produced in 
simulations of smuon pair production
for the supersymmetry parameter set considered in \cite{Blair}.  The technique 
generalizes to other supersymmetric particles.  The superpartner of the 
electron neutrino should often decay by 
\beq
           \s \nu \to   e^-  \s C^+_1 \ .
\eeq{snudecay}
The chargino decays to a complex final state, but the electron has the same 
flat distribution that we have just discussed.
Figure~\ref{fig:snu} shows a simulation study of the electron distribution
in $\s \nu$ pair-production, showing well-defined kinematic endpoints.
In chargino pair-production, the energy distribution is more complex, both
because the chargino decay is not isotropic and because the chargino decays
to a two-quark or two-lepton system of indefinite mass.   But the $q\bar q$
energy and mass distributions, shown in Fig.~\ref{fig:Cqqbar} still show 
quite well-defined endpoints and still allow very accurate mass
determinations~\cite{Blair}.

\begin{figure}[tb]
\begin{center}
\includegraphics[width=.6\textwidth]{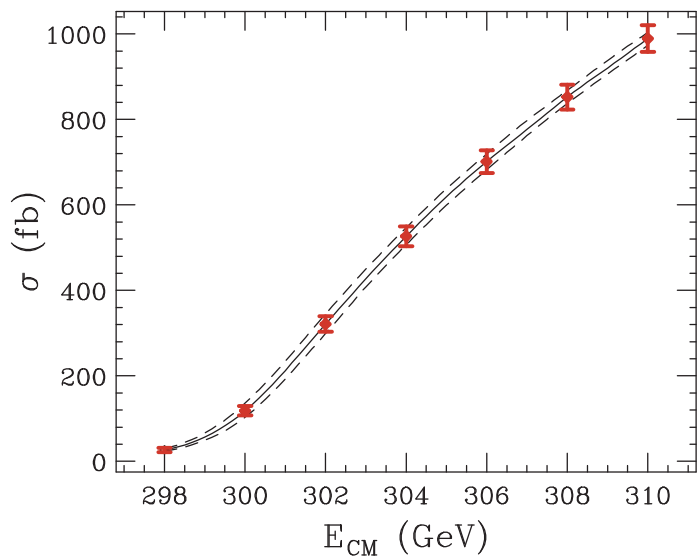}
\caption{Sensitivity of the threshold cross section in $e^-e^-\to \s{e}^-
\s{e}^-$ to the mass of the $\s e^-$,  from~\cite{Fengee}.  The three curves
correspond to selectron masses differing by 100 MeV.  Initial state 
radiation and other realistic beam effects are included.}
\label{fig:myeminus}
\end{center}
\end{figure}

The simplicity of these reactions can be further exploited along a number 
of lines
to expose more detailed aspects of supersymmetry spectroscopy.   Because it
is possible in $\ee$ annihilation to directly control the $\ee$ center of 
mass energy,
it is possible to precisely locate the threshold energy for 
a pair production
 process
\leqn{basicxxbar}.  This technique can produce a mass determination 
at the
0.1\% level.  In Fig.~\ref{fig:myeminus}, I show the dependence of the cross
 section
for the reaction $e^-e^- \to \s e^- \s e^-$ on center of mass energy in the
 vicinity 
of the threshold.  A variation of  the selectron mass by less than 0.1\% is
 quite
visible above the expected statistical errors~\cite{Fengee}. 
 
\begin{figure}[tb]
\begin{center}
\includegraphics[width=.43\textwidth]{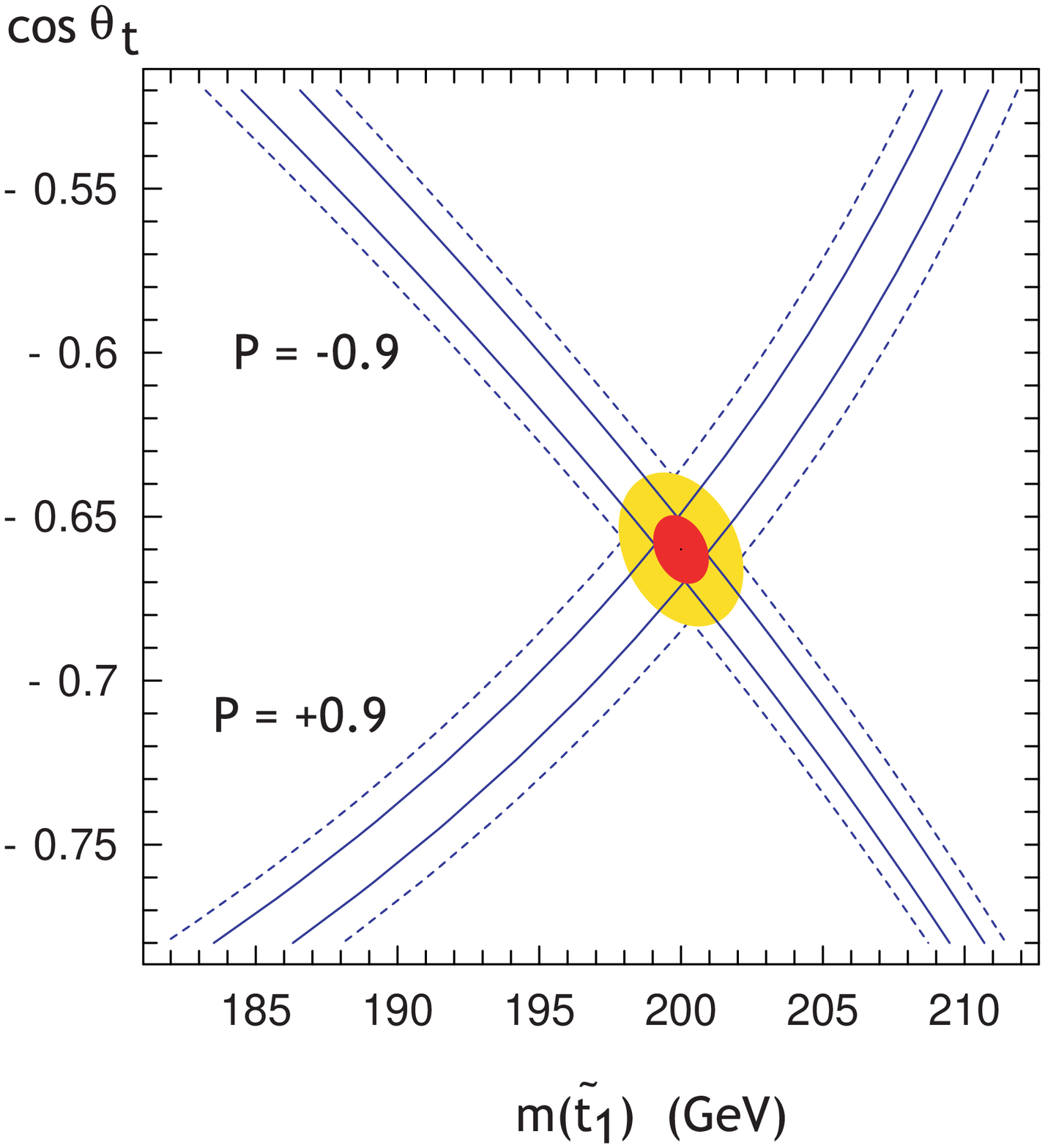}
\includegraphics[width=.54\textwidth]{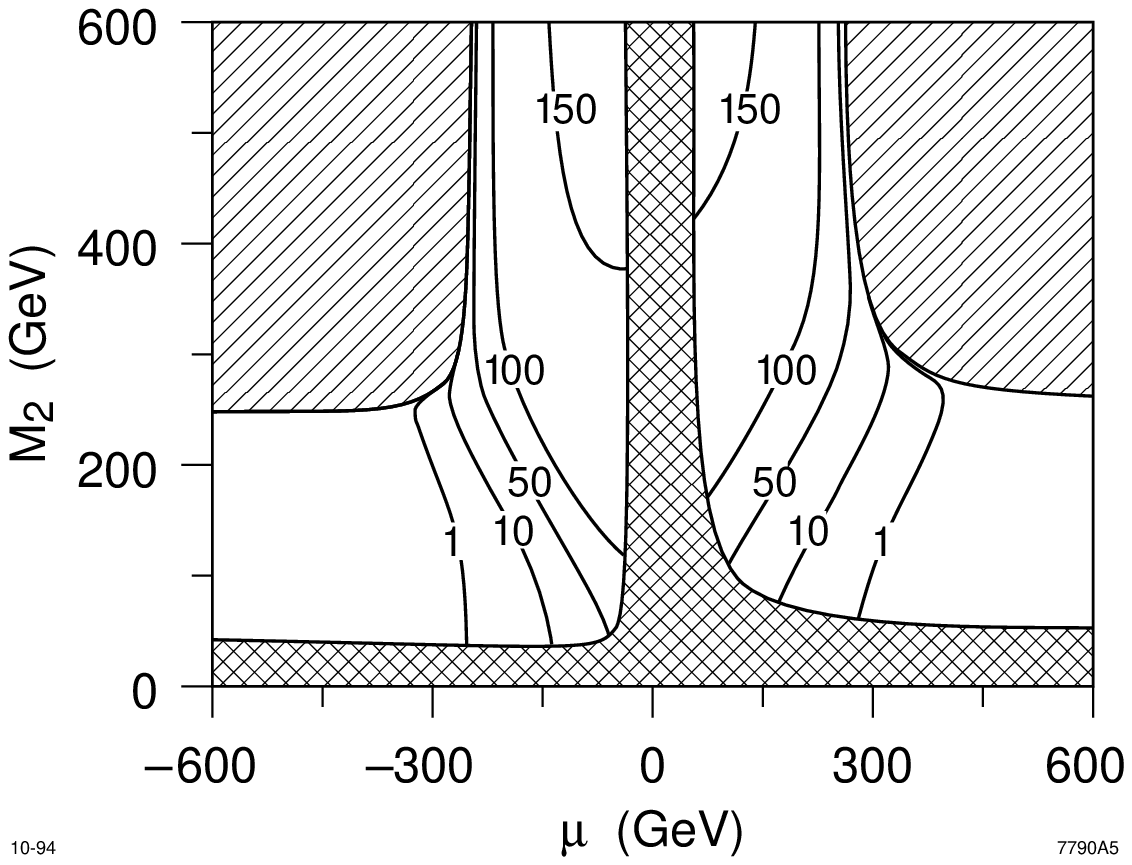}
\caption{Mixing angle determinations in $\ee$ annihilation to supersymmetric
 particles: Left: $\s t$ mixing angle determination from measurement of the
 pair-production cross section from left- and right-handed electron beams, 
 from \cite{Eberl}.
  Right:  $\mu$ vs. $m_2$ determination from measurement of the 
     production cross section for chargino pairs from using an $e^-_R$ beam, 
    from  \cite{FPMX}.}
\label{fig:stop}
\end{center}
\end{figure}

A more subtle question is the determination of the mixing angles defining
 the stop, stau, chargino, and neutralino eigenstates.  For this study, the 
initial electron polarization can be used in a powerful way.  For the stop and 
stau, the pair-production cross section for a given initial-state polarization
 depends only on the electroweak 
quantum numbers of the final particles.  The mass eigenstate is a mixture of 
two states with different quantum numbers, and so the cross section is 
an unambiguous function of the mixing angle.  Figure~\ref{fig:stop}(a) shows
a determination of the mixing angle in the lighter stop eigenstate by 
comparing 
the measured pair-production cross sections from left- and right-handed 
polarized beams.   For the charginos and neutralinos, the pair-production 
from 
left- and right-handed beams actually accesses different Feynman diagrams 
with
different intermediate particles.  For example, the production from a 
right-handed
electron beam (at least for center of mass energies much larger than $m_Z$)
produces only the component of the eigenstate that is the partner of the Higgs
boson.  Figure~\ref{fig:stop}(b) shows the value of this polarized production 
cross-section as a function of the parameters $\mu$ and $m_2$.  
The cross section is large is regions where the lightest 
chargino is mainly a Higgsino and 
small where it is mainly a gaugino.  The measured mass of the chargino  picks
out a specific point on each contour of constant cross section.  With this
constraint, the  content of the chargino 
eigenstate can be precisely determined.

\section{Conclusions}

In this lecture, I have presented a possible picture of the future of 
high-energy
physics based on the existence of supersymmetry, a fundamental symmetry
between bosonic and fermionic elementary particles.  After reviewing the 
current status of our understanding of the interactions of elementary 
particles,
I have explained how supersymmetry can address  many of the pressing questions
that are now unanswered.

But just as supersymmetry provides the solution to our present questions,
it will raise a new set of questions that must then be investigated.
Chief among these is the question of the mechanism of supersymmetry
breaking and the origin of the masses of superpartners.
I have argued that these questions might well connect directly to very deep 
issues of the short-distance geometry of spacetime and to the connection of 
the observed
interactions of particle physics to string theory or another grand theory of 
unification.

I have argued that these new questions will need to be resolved from 
experimental
data, specifically, the data on the masses and mixing of the new particles 
predicted
by supersymmetry.  I have explained how the next generation of 
particle accelerators will give us the tools to acquire this data. 
These are huge and 
expensive technical projects, but they have the capabilities to bring us the 
information that we need.

This experimental study will bring us into a new regime in fundamental physics,
and we must frankly acknowledge that we do not know what its outcome will 
be.  Perhaps the 
superspectrum measurements will show an anticipated, simple pattern.  More 
likely, as has happened for every other new set of particles and forces, 
they will present a puzzle that defies straightforward projections.  

This is what we hope for whenever we experiment on the laws of physics. 
We look for a chance
to raise puzzles whose resolution will take us deeper into the 
working of Nature.  To solve such puzzles, physicists must 
organize the facts into newly imagined patterns and regularities. 
Today the Standard Model leads us to the need for supersymmetric particles.
We look forward to their discovery, and then their painstaking exploration.
When the facts about these particles are gathered, we will find ourselves
with concrete questions that will challenge us to make another such leap.
We will find ourselves then at that moment that we prize, 
the moment when the  
next Werner Heisenberg can open our eyes to a yet more unexpected reality.

\bigskip
\bigskip
I am grateful to Professors Gerd Buschhorn
  and Julius Wess for their invitation to speak at
the symposium and  to Jonathan Feng, Michael Dine, Keisuke Fujii, Hitoshi 
Murayama, 
and many other colleagues 
at SLAC and elsewhere with whom I have discussed the issues presented in 
this lecture.  I thank Richard Zare for his very useful critique of  the
manuscript.  This work was supported by the US Department of Energy
under contract DE--AC03--76SF00515.


\begin{thebibliography}{99}

\bibitem{Rechenberg}
J. Mehra and H. Rechenberg, {\sl The Historical Development of Quantum 
Theory}, vol. 2,section V.5  (Springer-Verlag, New York, 1982).

 \bibitem{Uncertainty}
D. C. Cassidy, {\sl Uncertainty: the Life and Science of Werner Heisenberg}.
  (W. H. Freeman, New York, 1992).

\bibitem{YM}
C.~N.~Yang and R.~L.~Mills,
Phys.\ Rev.\  {\bf 96}, 191 (1954).



\bibitem{HM}
For an overview, see,
 \eg, F. Halzen and A. D. Martin, {\sl Quarks and Leptons}.
(Wiley, 1984).

\bibitem{OPALZ}
G.~Abbiendi {\it et al.}  [OPAL Collaboration],
Eur.\ Phys.\ J.\ C {\bf 19}, 587 (2001)
[arXiv: hep-ex/0012018].
 I thank T. Mori for permission to use
        this figure.

\bibitem{allZ}
 LEP Collaborations and the LEP Electroweak Working Group,
arXiv:hep-ex/0101027.

\bibitem{topmass}
B.~Abbott {\it et al.}  [D0 Collaboration],
Phys.\ Rev.\ D {\bf 58}, 052001 (1998)
[arXiv:hep-ex/9801025]; 
T.~Affolder {\it et al.}  [CDF Collaboration],
Phys.\ Rev.\ D {\bf 63}, 032003 (2001)
[arXiv:hep-ex/0006028].

\bibitem{LEPHiggs}
M.~W.~Grunewald,
arXiv:hep-ex/0210003, to appear in the Proceedings of the 31st Intl.
Conf. on High-Energy Physics, Amsterdam, 2002.

\bibitem{KadoTully}
M. M. Kado and C. G. Tully, Ann. Rev. Nucl. Part. Sci.
{\bf 52}, 65 (2002).

\bibitem{ErlerL}
J. Erler and P. Langacker, in K.~Hagiwara {\it et al.} 
 [Particle Data Group Collaboration],
Phys.\ Rev.\ D {\bf 66}, 010001 (2002).

\bibitem{Turner}
M. S. Turner, these proceedings; 
M.~S.~Turner,
Int.\ J.\ Mod.\ Phys.\ A {\bf 17}, 3446 (2002)
[arXiv:astro-ph/0202007].

\bibitem{GL}
Y.~A.~Golfand and E.~P.~Likhtman,
JETP Lett.\  {\bf 13} (1971) 323
[Pisma Zh.\ Eksp.\ Teor.\ Fiz.\  {\bf 13} (1971) 452].

\bibitem{VA}
D.~V.~Volkov and V.~P.~Akulov,
Phys.\ Lett.\ B {\bf 46}, 109 (1973).

\bibitem{WZ}
J.~Wess and B.~Zumino,
Nucl.\ Phys.\ B {\bf 70}, 39 (1974).

\bibitem{Nilles}
H.~P.~Nilles,
Phys.\ Rept.\  {\bf 110}, 1 (1984).

\bibitem{WessBagger}
J. Wess and J. Bagger, {\sl Supersymmetry and Supergravity}
 (Princeton Unversity Press, 1992).

\bibitem{Martin}
S.~P.~Martin,
in {\sl Perspectives on Supersymmetry}, G. L. Kane, ed. (World Scientific,
1998).
arXiv:hep-ph/9709356.

\bibitem{ColemanMandula}
S.~R.~Coleman and J.~Mandula,
Phys.\ Rev.\  {\bf 159}, 1251 (1967).

\bibitem{NS}
A.~Neveu and J.~H.~Schwarz,
Nucl.\ Phys.\ B {\bf 31}, 86 (1971).

\bibitem{Ramond}
P.~Ramond,
Phys.\ Rev.\ D {\bf 3}, 2415 (1971).

\bibitem{GSO}
F.~Gliozzi, J.~Scherk and D.~I.~Olive,
Nucl.\ Phys.\ B {\bf 122}, 253 (1977).

\bibitem{Polch}
J.~Polchinski, these proceedings,
arXiv:hep-th/0209105.

\bibitem{MYCERN}
M.~E.~Peskin, in {\sl Proceedings of the 1996 European School of High-Energy
Physics}, 
arXiv:hep-ph/9705479.

\bibitem{EllisRev}
J.~R.~Ellis, in {\sl Proceedings of the 1998 European School of High-Energy
Physics}, 
arXiv:hep-ph/9812235.

\bibitem{Schmaltz}
M.~Schmaltz, to appear in the  Proceedings of the 31st Intl.
Conf. on High-Energy Physics, Amsterdam, 2002.
arXiv:hep-ph/0210415.



\bibitem{neglect}
For the experts, the neglected effects are 2-loop renormalization group 
coefficients and high- and low-scale threshold corrections.  See, for example, 
P.~Langacker and N.~Polonsky,
Phys.\ Rev.\ D {\bf 52}, 3081 (1995)
[arXiv:hep-ph/9503214].

\bibitem{QEspin}
J.~R.~Espinosa and M.~Quiros,
Phys.\ Rev.\ Lett.\  {\bf 81}, 516 (1998)
[arXiv:hep-ph/9804235];
M.~Quiros and J.~R.~Espinosa,
arXiv:hep-ph/9809269.


\bibitem{ALEPHHiggs}
J.~A.~Kennedy  [ALEPH Collaboration],
arXiv:hep-ex/0111004.

\bibitem{OPALHiggs}
G.~Abbiendi {\it et al.}  [OPAL Collaboration],
arXiv:hep-ex/0209078.

\bibitem{Muong}
G.~W.~Bennett {\it et al.}  [Muon g-2 Collaboration],
Phys.\ Rev.\ Lett.\  {\bf 89}, 101804 (2002)
[Erratum-ibid.\  {\bf 89}, 129903 (2002)]
[arXiv:hep-ex/0208001].

\bibitem{Nyff}
M.~Knecht, A.~Nyffeler, M.~Perrottet and E.~De Rafael,
Phys.\ Rev.\ Lett.\  {\bf 88}, 071802 (2002)
[arXiv:hep-ph/0111059].


\bibitem{Davier}
M.~Davier, S.~Eidelman, A.~Hocker and Z.~Zhang,
arXiv:hep-ph/0208177.

\bibitem{Horava}
P.~Horava,
Phys.\ Rev.\ D {\bf 54}, 7561 (1996)
[arXiv:hep-th/9608019].


\bibitem{RandallSundrum}
L.~Randall and R.~Sundrum,
Nucl.\ Phys.\ B {\bf 557}, 79 (1999)
[arXiv:hep-th/9810155].

\bibitem{SchmaltzBr}
M.~Schmaltz and W.~Skiba,
Phys.\ Rev.\ D {\bf 62}, 095005 (2000)
[arXiv:hep-ph/0001172],
Phys.\ Rev.\ D {\bf 62}, 095004 (2000)
[arXiv:hep-ph/0004210].



\bibitem{ATLAS}
ATLAS Collaboration, {\sl Detector and Physics Performance Technical
     Design Report}, CERN/LHCC/99-14 (1999).

\bibitem{Graf}
I am grateful to N. Graf for providing this figure.

\bibitem{Tsukamoto}
T.~Tsukamoto, K.~Fujii, H.~Murayama, M.~Yamaguchi and Y.~Okada,
Phys.\ Rev.\ D {\bf 51}, 3153 (1995).

\bibitem{Blair}
H.~U.~Martyn and G.~A.~Blair, in {\sl Physics and Experiments with
Future Linear $\ee$ Colliders}, E. Fernandez and A. Pacheco, eds.
(Univ. Auton. de Barcelona, 2000).
arXiv:hep-ph/9910416.

\bibitem{Baer}
S.~Kuhlman {\it et al.},  [NLC ZDR Design Group and NLC Physics Working Group
                  Collaboration],
{\sl Physics and technology of the Next Linear Collider: 
A Report submitted to Snowmass '96},
arXiv:hep-ex/9605011.


\bibitem{Fengee}
J.~L.~Feng and M.~E.~Peskin,
Phys.\ Rev.\ D {\bf 64}, 115002 (2001)
[arXiv:hep-ph/0105100].

\bibitem{Eberl}
H.~Eberl, S.~Kraml, W.~Majerotto, A.~Bartl and W.~Porod,
 in {\sl Physics and Experiments with
Future Linear $\ee$ Colliders}, E. Fernandez and A. Pacheco, eds.
(Univ. Auton. de Barcelona, 2000).
arXiv:hep-ph/9909378.

\bibitem{FPMX}
J.~L.~Feng, M.~E.~Peskin, H.~Murayama and X.~Tata,
Phys.\ Rev.\ D {\bf 52}, 1418 (1995)
[arXiv:hep-ph/9502260].

\end{thebibliography}
\end{document}